\documentclass[showpacs,nofootinbib]{revtex4}
\usepackage{amsfonts}
\usepackage{amsmath,amssymb}
\usepackage{epsfig,graphicx}
\usepackage{dcolumn}

\input{tcilatex}
\begin{document}

\title{Poincar\'{e} gauge gravity: An emergent scenario}
\author{J.L. Chkareuli$^{1,2}$}
\affiliation{$^{1}$Center for Elementary Particle Physics, ITP, Ilia State University,
0162 Tbilisi, Georgia \\
$^{2}$Andronikashvili Institute of Physics, 0177 Tbilisi, Georgia}
\date{\today }

\begin{abstract}
The Poincar\'{e} gauge gravity (PGG) with the underlying vector fields of
tetrads and spin-connections is perhaps the best theory candidate for
gravitation to be unified with the other three elementary forces of nature.
There is a clear analogy between local frame in PGG and local internal
symmetry space in the Standard Model. As a result, the spin-connection
fields, gauging the local frame Lorentz symmetry group $SO(1,3)_{LF}$,
appear in PGG much as photons and gluons appear in SM. We propose that such
an analogy may follow from their common emergent nature allowing to derive
PGG in the same way as conventional gauge theories. In essence, we start
with an arbitrary theory of some vector and fermion fields which possesses
only global spacetime symmetries, such as Lorentz and translational
invariance, in flat Minkowski space. The two vector field multiplets
involved are proposed to belong, respectively, to the adjoint ($A_{\mu
}^{ij} $) and vector ($e_{\mu }^{i}$) representations of the starting global
Lorentz symmetry. We show that if these prototype vector fields are
covariantly constrained, $A_{\mu }^{ij}A_{ij}^{\mu }=\pm M_{A}^{2}$ and $%
e_{\mu }^{i}e_{i}^{\mu }=\pm M_{e}^{2}$, thus causing a spontaneous
violation of the accompanying global symmetries ($M_{A,e}$ are their
proposed violation scales), then the only possible theory compatible with
these length-preserving constraints is turned out to be the gauge invariant
PGG, while the corresponding massless (pseudo)Goldstone modes are naturally
collected in the emergent gauge fields of tetrads and spin-connections. In a
minimal theory case being linear in a curvature we unavoidably come to the
Einstein-Cartan theory. The extended theories with propagating
spin-connection and tetrad modes are also considered and their possible
unification with the Standard Model is briefly discussed.
\end{abstract}

\maketitle

\section{Introduction}

A conventional view on the problem of unification of all fundamental
(electromagnetic, weak and strong) interactions with gravity is based on the
modern superstring theory. However, this approach, as is well known, has
serious problems to be adapted to the present particle physics
phenomenology. Actually, only a few qualitative prescriptions stemming from
superstrings can really be taken into account at lower energies. In this
connection, one may think that a better understanding of some aspects of
this problem is still related to the quantum field theory framework provided
that the gravity, like as the other basic forces, is given by a true gauge
theory.

\subsection{PGG: an introductory commentary}

The general relativity (GR)\ is, of course, a gauge theory in a sense that
it is invariant under local general linear transformation group $GL(4)$
being considered according to the principle of general covariance as a
linear part of some fundamental world symmetry for our spacetime\footnote{%
This is certainly the most natural choice for a gauge group in GR if one
works in an ordinary metric formulation.}. The point is, however, that it is
not a theory which is similar to other gauge theories incorporated into the
Standard Model (SM) where gauge fields are in fact propagating physical
fields like photons, gluons and weak bosons being related to generators of
internal symmetry groups. In this connection, a natural question arises
whether a similar view could be applied to local spacetime symmetries that
could lead to a deep conceptual analogy between gravity and the other three
elementary forces described by modern gauge theories.

Indeed, GR can be described by a simple extension of the methods used to
describe Yang-Mills theory \cite{Yang-Mills}, as was argued by Utiyama \cite%
{Utiyama} a long ago. Traditionally, one might start with coordinate
transformations being a local generalization of translations, since gravity
is generically defined as a force that couples to energy-momentum in much
the same way as electromagnetism couples to charge. However, this is not
enough to define spinors, since there no representations of $GL(4)$ which
behave like spinors under its Lorentz subgroup. So, one must introduce some
extra local Lorentz symmetry to include spinors into the theory. Remarkably,
regardless of spinors, this is also required by the Einstein equivalence
principle for the local spacetime structure being identified with Minkowski
space. One might then propose \cite{Utiyama} that a true theory of gravity
can be seen as a gauge theory based on the local Lorentz group $SO(1,3)$ in
the same manner as the Yang-Mills gauge theory appears on the basis of the
internal isospin gauge group $SU(2)$. In this formulation, the affine
connection in a local Lorentz frame could be obtained as the gravitational
counterpart of the Yang-Mills gauge fields. However, this seems to be only
half a story. In order to relate this local Lorentz frame to the global
world space (being curved in general), one needs some extra tools. They are
tetrad fields being built into all spacetime points to define local
reference frames. Tetrads may be considered as objects given \textit{a
priori }\cite{Utiyama} or one can try to treat them on the gauge base as
well relating them to the local translations in a general frame \cite{Kibble}%
. So, the gauge fields in the complete Poincar\'{e} gauge gravity (PGG)
include the tetrads $e_{\mu }^{i}$ as well as the local affine connections $%
A_{\mu }^{ij}$ which are usually referred to as the spin-connections in PGG.
Remarkably, though PGG\ can be thought as a field theory in flat Minkowski
space, it successfully mimics curved space geometry when is expressed in
terms of the general affine connection and metric. Consequently, the local
Poincar\'{e} symmetry appears to correspond to a world space with torsion as
well as curvature: just torsion is a feature with which the base space
should be endowed to accommodate spinor fields. In this sense, PGG has the
geometric structure of the Riemann--Cartan space $U_{4}$.

We have, therefore, in PGG the world coordinate transformation group, which
includes translations and the orbital part of Lorentz transformations acting
in the global base space, and a locally acting Lorentz group, which includes
the spin part of Lorentz transformations in the flat Minkowski spacetime. To
make a distinction between them we use Greek indices $(\mu $, $\nu ,\rho
,...=0$, $1$, $2$, $3)$ for the former and Latin indices $%
(a,b,c,...,i,j,k,...=0$, $1$, $2$, $3)$ for the latter. All field variables
taken in the theory, including tetrads and spin-connections, are generally
proposed to be functions of the global world 4-coordinate $x_{\mu }$.
Clearly, this coordinate itself and therefore its partial derivatives $%
\partial _{\mu }$, are not affected by the spin part of Lorentz
transformations, while all spinor and vector fields are acted on. {}In
essence, the vector gauge fields of tetrads $e^{i}{_{\mu }}$ ($e^{\mu }{_{i})%
}$ satisfying the following orthogonality conditions 
\begin{equation}
e_{\mu }^{i}e_{i}^{\nu }\,=\delta _{\mu }^{\nu }\,\text{,}~~~e_{\mu
}^{i}e_{j}^{\mu }\,=\delta _{j}^{i}  \label{t}
\end{equation}%
set the local coordinate frame, while those of the spin-connections $A_{\mu
}^{ij}$ are in fact connections with respect to the tetrad frame.
Remarkably, this "global-local" duality is in a natural accordance with the
Einstein equivalence principle which is\ somewhat hidden in the standard GR
and should be stated per se.

An important point\ to stress is that the Poincar\'{e} gauge theory of
gravitation is basically related to the vector fields of tetrads and
spin-connections as the primary notions in theory. This is quite
understandable since gauge field is a tool which extends an ordinary
derivative to the covariant one. Thus, such a field may be only a vector
field, by definition. In this sense, an effective tensor field of metric 
\begin{equation}
g_{\mu \nu }=e_{\mu }^{i}e_{\nu }^{j}\eta _{ij}  \label{metric}
\end{equation}%
(where $\eta _{ij}$ is the metric in the Minkowski spacetime, $\eta
_{00}=-\eta _{11}=-\eta _{22}=-\eta _{33}=1$) being in GR traditionally
identified with a graviton, appears as an essentially secondary tool in this
scheme. In this connection, PGG is in principle different theory which may
have some common root with conventional gauge theories, as can also be
clearly seen from its\ emergent scenario presented below. In any case, it
certainly is a valuable alternative to the standard GR and has extra
opportunities to describe gravity in a self-consistent way [4-8]. Some
excellent presentations can be found in the reviews [9-13].

\subsection{The present paper}

All the above allows to think that PGG with the underlying vector fields of
tetrads and spin-connections is perhaps the best theory candidate for
gravitation to be unified with the other three elementary forces in the
quantum field theory framework. Remarkably, there is some clear analogy
between a local frame in PGG and a local internal symmetry space in
conventional quantum field theories. As a result, the vector fields of the
spin-connections $A_{\mu }^{ij}(x)$ gauging the local frame Lorentz group $%
SO(1,3)_{LF}$ as its adjoint representation, appear in PGG much as photons
and gluons appear in in the Standard Model. We propose that such an analogy
may follow from their common emergent nature allowing to derive PGG in the
same way as conventional gauge theories. To make this analogy clearer we
give in the next Section II a detailed presentation of the emergent gauge
theories. We begin with an underlying emergence conjecture \cite{cfn2, c}
according to which the vector field theory possessing only some input global
internal symmetry may convert this symmetry into the local one provided that
the vector field (or vector field multiplet) involved is covariantly
constrained, $A_{\mu }A^{\mu }=\pm M^{2}$ (where $M$ is some mass scale).
Normally, this constraint might mean a spontaneous Lorentz violation with
the vector Goldstone bosons produced, while the vector Higgs component
appears frozen to its vacuum expectation value (VEV) $M$. This is what
usually happens in the non-linear $\sigma $-model for pions \cite{GLA} with
a spontaneously broken chiral symmetry. However, in the vector field
theories due to gauge invariance emerged the physical Lorentz invariance is
left intact. Remarkably, the emergence conjecture may be applied to any
vector field theory with an input global Abelian or non-Abelian internal
symmetry and, consequently, the conventional gauge invariant QED or
Yang-Mills theory are emerged. This is explicitly demonstrated at some
length in the Sections II A and II B, respectively.

In the Section III we turn to the construction of an emergent PGG theory. We
start with an arbitrary theory of some vector and fermion fields which
possesses only global spacetime symmetries, such as Lorentz and
translational invariance, in flat Minkowski space $M_{4}$. The two vector
field multiplets involved are proposed to belong, respectively, to the
adjoint ($A_{\mu }^{ij}$) and vector ($e_{\mu }^{i}$) representations of the
starting global Lorentz symmetry. We show that if these prototype vector
fields are covariantly constrained, $A_{\mu }^{ij}A_{ij}^{\mu }=\pm M_{A}^{2}
$ and $e_{\mu }^{i}e_{i}^{\mu }=\pm M_{e}^{2}$, thus causing a spontaneous
violation of the accompanying global symmetries ($M_{A,e}$ are their
proposed violation scales), then the only possible theory compatible with
these constraints is turned out to be the standard PGG (with gauged
translations and local frame Lorentz transformations), while the
corresponding massless (pseudo)Goldstone modes are naturally collected in
the emergent gauge fields of tetrads and spin-connections. In essence, the
local Poincar\'{e} invariance emerges as a necessary condition for the
prototype vector fields $A_{\mu }^{ij}$ and $e_{\mu }^{i}$ not to be
superfluously restricted in degrees of freedom, apart from the above
length-fixing constraints. Again, due to gauge invariance emerged the
physical Lorentz and translation invariance remains in the final theory. In
minimal theory case being linear in curvature we unavoidably come to the
Einstein-Cartan theory that is thoroughly presented in the Section IV. The
extended theories with propagating spin-connection and tetrad modes are also
considered and their possible unification with the Standard Model is briefly
discussed. And in the final Section V we conclude.

\section{Emergent gauge theories}

The dynamical origin of massless particle excitations for spontaneously
broken internal symmetries \cite{NJL} allows to think that possibly only
global symmetries are fundamental symmetries of nature, whereas local
symmetries and associated massless gauge fields could solely emerge due to
spontaneous breaking of underlying spacetime symmetries involved, such as
relativistic invariance. It is conceivable that spontaneous Lorentz
invariance violation (SLIV) could provide a dynamical approach to quantum
electrodynamics, gravity and Yang-Mills theories with the photon, graviton
and gluons appearing as massless Nambu-Goldstone bosons. This rather old
idea \cite{bjorken,ph,eg,suz} has gained a further development \cite%
{cfn,kraus,kos,car,cjt} in recent years.

We will follow here a somewhat different point of view \cite{cfn2, c}
according to which SLIV in itself could be only one of the consequences of
the gauge symmetry emergence scenario rather than its underlying cause.
Presumably, such a scenario is basically related to some covariant
constraint(s) which, for one reason or another, is put on a vector field
system possessing only some global internal symmetry. As a matter of fact,
the simplest holonomic constraint of this type for vector field (or vector
field multiplet) $A_{\mu }$\ may be the "length-fixing" condition 
\begin{equation}
C(A)=A_{\mu }A^{\mu }-n^{2}M^{2}=0\text{ , \ \ }n^{2}\equiv n_{\mu }n^{\mu
}=\pm 1\text{ }  \label{const}
\end{equation}%
where $n_{\mu }$ is a properly oriented unit Lorentz vector, while $M$ is
some high mass scale. We will see that gauge invariance appears unavoidable
in the proposed theory, if the equations of motion involved should have
enough freedom to allow a constraint like (\ref{const}) to be fulfilled and
preserved over time. Namely, gauge invariance in such theories has to appear
in essence as a response of an interacting field system\textit{\ }to putting
the covariant constraint (\ref{const}) on its dynamics, provided that we
allow parameters in the corresponding Lagrangian density to be adjusted so
as to ensure self-consistency without losing too many degrees of freedom.
Otherwise, a given field system could get unphysical in a sense that a
superfluous reduction in the number of degrees of freedom would make it
impossible to set the required initial conditions in an appropriate Cauchy
problem. Namely, it would be impossible to specify arbitrarily the initial
values of the vector and other field components involved, as well as the
initial values of the momenta conjugated to them. Furthermore, in quantum
theory, to choose self-consistent equal-time commutation relations would
also become impossible \cite{ogi3}.

Let us dwell upon this point in more detail. Generally, while a conventional
variation principle requires the equations of motion to be satisfied, it is
possible to eliminate one spacetime component of a general 4-vector field $%
A_{\mu }$ in order to describe a pure spin-1 particle by imposing a
supplementary condition. In the massive vector field case there are three
physical spin-1 states to be described by the $A_{\mu }$ field. Similarly in
the massless vector field case, although there are only two physical
(transverse) photon spin states, one cannot construct a massless 4-vector
field $A_{\mu }$ as a linear combination of creation and annihilation
operators for helicity $\pm 1$ states in the Lorentz covariant way, unless
one fictitious state is added \cite{GLB}. So, in both the massive and
massless vector field cases, only one component of the $A_{\mu }$ field may
be eliminated and still preserve physical Lorentz invariance. Now, once the
constraint (\ref{const}) is imposed, it is therefore not possible to satisfy
another supplementary condition since this would superfluously restrict the
number of degrees of freedom for the vector field. To avoid this, its
equation of motion should not lead by itself to any new constraint that is
only possible if it is automatically divergenceless, as normally appears in
the gauge invariant theory. Thus, due the constraint (\ref{const}), being
the only possible covariant and holonomic vector field constraint, the
theory has to acquire on its own a gauge-type invariance, which gauges the
starting global symmetry of the interacting vector and matter fields
involved.

\subsection{A QED primer}

To see how technically a global internal symmetry may be converted into a
local one in an Abelian $U(1)$ symmetry case, let us consider in detail the
question of consistency of the constraint for vector field (\ref{const})
with its equations of motion in an arbitrary relativistically invariant
Lagrangian $L(A,\psi )$ which also contains some charged fermion field $\psi 
$. In the presence of the constraint (\ref{const}), it follows that the
equations of motion can no longer be independent. The important point is
that, in general, the time development would not preserve the constraint.
So, the parameters in the Lagrangian have to be chosen in such a way that
effectively we have one less equation of motion for the vector field. This
means that there should be some relationship between all the vector and
matter field Eulerians ($E_{A}$, $E_{\psi }$, ...) involved\footnote{%
Hereafter, the notation $E_{A}$ stands for the vector field Eulerian $%
(E_{A})^{\mu }\equiv \partial L/\partial A_{\mu }-\partial _{\nu }[\partial
L/\partial (\partial _{\nu }A_{\mu })].$ We use similar notations for other
field Eulerians as well.}. Such a relationship can quite generally be
formulated as a functional - but by locality just a function - of the
Eulerians, $F(E_{A},E_{\psi })$, being put equal to zero at each spacetime
point with the configuration space restricted by the constraint $C(A)=0$, 
\begin{equation}
F(C=0;\text{ \ }E_{A},E_{\psi })=0\text{ .}  \label{FF}
\end{equation}%
for the one matter fermion case proposed. This relationship, which we shall
call the emergence equation for what follows, must satisfy the same symmetry
requirements of Lorentz and translational invariance, as well as all the
global internal symmetry requirements, as the general starting Lagrangian
does. This Lagrangian is supposed to also include the standard Lagrange
multiplier term with the field $\lambda (x)$ 
\begin{equation}
L^{tot}(A,\psi ,\lambda )=L(A,\psi )-\frac{\lambda }{2}\left( A_{\mu }A^{\mu
}-n^{2}M^{2}\right)  \label{44}
\end{equation}%
the variation under which results in the above constraint $C(A)=0$. In fact,
the relationship (\ref{FF}) is used as the basis for an emergence of gauge
symmetries in the constrained vector field theories \cite{cfn2, c}. We
propose that initial values for all fields (and their momenta) involved are
chosen so as to restrict the phase space to values with a vanishing
multiplier function $\lambda (x)$, $\lambda $ $=0$. Actually, due to an
automatic conservation of the Noether matter current in the theory an
initial value $\lambda =0$ will then remain for all time so that the
Lagrange multiplier field $\lambda $ never enters in the physical equations
of motions for what follows\footnote{%
A more direct way to have this solution with the vanished Lagrange
multiplier field $\lambda (x)$ would be to include in the Lagrangian (\ref%
{44}) an additional Lagrange multiplier term of the type $\xi \lambda ^{2}$,
where $\xi (x)$ is a new multiplier field. One can then easily confirm that
a variation of the modified Lagrangia $L^{\prime }+$ $\xi \lambda ^{2}$ with
respect to the $\xi $ field leads to the condition $\lambda =0$, whereas a
variation with respect to the basic multiplier field $\lambda $ preserves
the vector field constraint (\ref{const}). Such an approach with additional
Lagrange multiplier terms can be then properly generalized to the emergent
Yang-Mills theory and PGG theory cases considered later.}.

Let us consider a \textquotedblleft Taylor expansion" of the function $F$
expressed through various combinations of the fields involved, their
combinations with the Eulerians, as well as the derivatives acting on them.
The constant term in this expansion is of course zero since the relation (%
\ref{FF}) must be trivially satisfied when all the Eulerians vanish, i.e.
when the equations of motion are satisfied. We basically consider the terms
with the lowest mass dimension 4, corresponding to the Lorentz invariant
expressions 
\begin{equation}
\partial _{\mu }(E_{A})^{\mu },\text{ }A_{\mu }(E_{A})^{\mu },\text{ }%
E_{\psi }\psi ,\text{ }\overline{\psi }E_{\overline{\psi }}.  \label{fff}
\end{equation}%
to eventually have an emergent gauge theory at a renormalizible level. All
the other terms in the expansion contain field combinations and derivatives
with higher mass dimension and must therefore have coefficients with an
inverse mass dimension. We expect the mass scale associated with these
coefficients should correspond to a large fundamental mass (e.g. the Planck
mass $M_{P}$). Hence we conclude that such higher dimensional terms must be
properly suppressed and can be neglected.

Now, under the assumption that the constraint (\ref{const}) is preserved
under the time development given by the equations of motion, we show how
gauge invariance of the starting Lagrangian $L(A,\psi )$ in (\ref{44}) is
established. A conventional variation principle applied to the total
Lagrangian\ $L^{tot}(A,\psi ,\lambda )$ requires the following equations of
motion for the vector field $A_{\mu }$ and the auxiliary field $\lambda $ to
be satisfied%
\begin{equation}
(E_{A})^{\mu }=0\text{ , \ \ \ }C(A)=A_{\mu }A^{\mu }-n^{2}M^{2}=0\text{\ }.
\end{equation}%
However, in accordance with general arguments given above, the existence of
five equations for the 4-component vector field $A^{\mu }$ (one of which is
the constraint) means that not all of the vector field Eulerian components
can be independent. Therefore, there must be a relationship of the form
given in the emergence equation (\ref{FF}). When being expressed as a linear
combination of the Lorentz invariant terms (\ref{fff}), this equation leads
to the identity between the vector and matter field Eulerians of the
following type 
\begin{equation}
\partial _{\mu }(E_{A})^{\mu }=itE_{\psi }\psi -it\overline{\psi }E_{%
\overline{\psi }}.  \label{div}
\end{equation}%
(where $t$ is some constant) which is in fact identically vanished when the
equations of motion are satisfied\footnote{%
Note the term proportional to the vector field itself, $A_{\mu }(E_{A})^{\mu
}$, which would correspond to the selfinteraction of vector field, is absent
in the identity (\ref{div}). In presence of this term the transformations of
the vector field given below in (\ref{trans}) would be changed to $\delta
A_{\mu }=\partial _{\mu }\omega +c\omega A_{\mu }$. The point is, however,
that these transformations cannot in general form a group unless the
constant $c$ vanishes, as can be readily confirmed by constructing the
corresponding Lie bracket operation for two successive vector field
variations. We shall see later that non-zero $c$-type coefficients
necessarilly appear in the non-Abelian internal symmetry case, resulting
eventually in a emergent gauge invariant Yang-Mills theory.}. This identity
immediately signals about invariance of the basic Lagrangian $L(A,\psi )$ in
(\ref{44}) under vector and fermion field local $U(1)$ transformations whose
infinitesimal form is given by 
\begin{equation}
\delta A_{\mu }=\partial _{\mu }\omega ,\text{ \ \ }\delta \psi \text{\ }%
=it\omega \psi \text{ .}  \label{trans}
\end{equation}%
Here $\omega (x)$ is an arbitrary function, only being restricted by the
requirement to conform with the nonlinear constraint (\ref{const}) 
\begin{equation}
(A_{\mu }+\partial _{\mu }\omega )(A^{\mu }+\partial ^{\mu }\omega
)=n^{2}M^{2}\text{ .}  \label{gau}
\end{equation}%
Conversely, the identity (\ref{div}) follows from the invariance of the
physical Lagrangian $L(A,\psi )$ under the transformations (\ref{trans}).
Indeed, both direct and converse assertions are particular cases of
Noether's second theorem \cite{noeth}.

So, we have shown how the constraint (\ref{const}) enforces the choice of
the parameters in the starting Lagrangian $L^{tot}(A,\psi ,\lambda )$, so as
to convert its global $U(1)$ charge symmetry into a local one, thus
demonstrating an emergence of gauge symmetry (\ref{trans}) that allows the
emerged Lagrangian to be completely determined. For a theory with
renormalizable couplings, it is in fact the conventional QED Lagrangian

\begin{equation}
L_{QED}^{em}(A,\psi )=-\frac{1}{4}F_{\mu \nu }F^{\mu \nu }+\overline{\psi }%
(i\gamma \partial +m)\psi -eA_{\mu }\overline{\psi }\gamma ^{\mu }\psi \text{
\ }  \label{lag11}
\end{equation}%
extended by the Lagrange multiplier term, which provides the constraint (\ref%
{const}) imposed on the vector field $A_{\mu }.$ Interestingly, this type of
the QED theory with the constrained vector potential was considered by Nambu 
\cite{nambu} quite a long ago.

Let us make it clearer what does the constraint (\ref{const}) mean in the
gauge invariant QED framework. This constraint is in fact very similar to
the constraint appearing in the nonlinear $\sigma $-model for pions \cite%
{GLA}. It means, in essence, that the vector field $A_{\mu }$ develops some
constant background value, $\left\langle A_{\mu }\right\rangle =n_{\mu }M$,
and the Lorentz symmetry $SO(1,3)$ formally breaks down to $SO(3)$ or $%
SO(1,2)$ depending on the time-like ($n^{2}=1$) or space-like ($n^{2}=-1$)
nature of SLIV with the corresponding vector Goldstone mode associated with
a photon. Nonetheless, despite an evident similarity with the nonlinear $%
\sigma $-model for pions, which really breaks the corresponding chiral $%
SU(2)\times SU(2)$ symmetry in hadron physics, the QED theory (\ref{lag11})
with the supplementary vector field constraint (\ref{const}) involved leaves
the physical Lorentz invariance intact.

Some heuristic argument may look as follows. Let us turn to the convenient
SLIV\ parametrization%
\begin{eqnarray}
A_{\mu } &=&a_{\mu }+n_{\mu }H\text{ , \ }n_{\mu }a^{\mu }=0\text{ \ \ } 
\notag \\
\text{\ \ }H &=&(M^{2}-n^{2}a^{2})^{1/2}=M-\frac{a^{2}}{2M}+\cdot \cdot
\cdot \text{ \ (}a^{2}\equiv a_{\mu }a^{\mu }\text{)}  \label{ab}
\end{eqnarray}%
which could be referred to as the "symmetry broken phase" being determined
by the constraint (\ref{const}) itself. Indeed, the new $a^{\mu }$ field
appears here as the vector\ Goldstone mode which is orthogonal to the vacuum
direction given by the unit vector $n_{\mu }$, while $H$ stands for the
effective Higgs mode. Substituting the parametrization (\ref{ab}) into the
emergent Lagrangian (\ref{lag11}) and properly redefining the fermion field
with the phase linear in coordinate 
\begin{equation}
\psi \rightarrow e^{ieM(x\cdot n)}\psi  \label{red}
\end{equation}%
to exclude the fictitious noncovariant mass term $eM\overline{\psi }n_{\mu
}\gamma ^{\mu }\psi $ one comes to the theory expressed solely in the vector
Goldstone modes $a_{\mu }$ emerging in the "axial gauge" (\ref{ab})%
\begin{eqnarray}
L(a,\psi ) &=&-\frac{1}{4}[f_{\mu \nu }+(\partial _{\mu }n_{\nu }-\partial
_{\nu }n_{\mu })H][f^{\mu \nu }+(n^{\mu }\partial ^{\nu }-n^{\nu }\partial
^{\mu })H]-\frac{1}{2}\delta (n_{\mu }a^{\mu })^{2}  \notag \\
&&+\overline{\psi }(i\gamma \partial +m)\psi -ea_{\mu }\overline{\psi }%
\gamma ^{\mu }\psi +(H-M)\overline{\psi }n_{\mu }\gamma ^{\mu }\psi .
\label{NL}
\end{eqnarray}%
where we have denoted the Goldstone field strength tensor by $f_{\mu \nu
}=\partial _{\mu }a_{\nu }-\partial _{\nu }a_{\mu }$ and also retained the
notation for the redefined fermion field. Now, one can see that though the
Lagrangian (\ref{NL}) contains a plethora of Lorentz violating couplings
(stemming from the effective Higgs field expansion (\ref{ab}) in it) they
all disappear in the limit $M\rightarrow \infty $ that leads to the theory
which is completely equivalent to a conventional QED taken in an axial
gauge. Thus, a possible Lorentz violation, even if it were the case, should
only be extremely small being significantly suppressed by inverse powers of
the large mass scale $M$. However, as was shown in the tree \cite{nambu} and
one-loop \cite{az} approximations, there is no physical Lorentz violation in
the emergent QED for any mass value $M$ in the vector field constraint (\ref%
{const}). Later this result was also confirmed for many other gauge theories
with the supplementary vector field constraints, such as the spontaneously
broken massive QED \cite{kep}, non-Abelian theory \cite{jej, cj}, tensor
field gravity \cite{cjt} and also their supersymmetric extensions \cite{c}.

So, in contrast to a spontaneous violation of internal symmetries, SLIV
caused by the length-preserving constraint does not necessarily imply a
physical breakdown of Lorentz invariance. Rather, we are concerned here only
with a "spontaneous breakdown" of an input covariance of the gauge condition
(\ref{const}) for the starting vector field $A_{\mu }$ to the noncovariant
axial gauge (\ref{ab}) for the appearing vector Goldstone boson $a_{\mu }$.
In this connection, though it may seem counterintuitive, these two aspects
could be allowed to coexist: an appearance of vector Goldstone bosons, from
the one side, and a non-observability of spontaneous Lorentz violation
caused by a covariant constraint, from the other. Actually, the constrained
gauge fields are turned out to be, at the same time, the vector Goldstone
bosons manifesting themselves in the "symmetry broken phase" (\ref{ab}).
However, gauge invariance in QED and other gauge theories always leads to a
total conversion of SLIV into gauge degrees of freedom of massless vector
Goldstone bosons. We will hereafter refer to this case of SLIV as an
"inactive" SLIV, as opposed to an "active" SLIV leading to physical Lorentz
violation which appears if gauge invariance in the theory is explicitly
broken\footnote{%
Indeed, the only way for an inactive SLIV to to be activated (thus causing a
physical Lorentz violation) would appear only if gauge invariance in the
theory were really broken rather than merely constrained by some gauge
condition. Such a violation of gauge invariance could provide some extension
of the considered model with high-dimension operators induced presumably by
quantum gravity at very small distances \cite{par}.}.

\subsection{Yang-Mills theories}

We still have considered only a single vector field case with an underlying
global $U(1)$ symmetry. An extension to a theory possessing from the outset
some global non-Abelian symmetry $G$ follows the same logic, though includes
some peculiarities \cite{cfn2, c}. Suppose that this theory contains an
adjoint vector field multiplet \textbf{\ }$\boldsymbol{A}_{\mu }^{p}$ and
some fermion matter field multiplet $\boldsymbol{\psi }$ \ belonging to one
of irreducible representations of $G$ given by matrices $t^{p}$%
\begin{equation}
\lbrack t^{p},t^{q}]=if^{pqr}t^{r}\text{ , \ }Tr(t^{p}t^{q})=\delta ^{pq}%
\text{ }\ \text{(}p,q,r=0,1,...,N-1\text{)\ }  \label{22}
\end{equation}%
where $f^{pqr}$ stand structure constants, while $N$ is a dimension of the $%
G $ group. The corresponding Lagrangian \textrm{L}$^{tot}$ is supposed to
also include the standard Lagrange multiplier term with the field function $%
\mathrm{\lambda }(x)$ 
\begin{equation}
\mathrm{L}^{tot}(\boldsymbol{A}_{\mu },\boldsymbol{\psi },\text{ }\mathrm{%
\lambda })=\mathrm{L}(\boldsymbol{A}_{\mu },\boldsymbol{\psi })-\text{ }%
\frac{\mathrm{\lambda }}{2}(\boldsymbol{A}_{\mu }^{p}\boldsymbol{A}^{p\mu }-%
\boldsymbol{n}^{2}\mathrm{M}^{2})\text{ , \ }\boldsymbol{n}^{2}\equiv 
\boldsymbol{n}_{\mu }^{p}\boldsymbol{n}^{p\mu }=\pm 1  \label{33}
\end{equation}%
where $\boldsymbol{n}_{\mu }^{p}$ stands now for some properly-oriented
`unit' rectangular matrix both in Minkowski spacetime and internal space. As
in the above Abelian case, the Lagrange multiplier term does not contribute
to the vector field equations of motion by the proper choice of initial
values for all fields (and their momenta) involved so as to restrict the
phase space to values with a vanishing multiplier function $\mathrm{\lambda }%
(x)$ (see also the footnote$^{3}$).

The variation of the Lagrangian $\mathrm{L}^{tot}(\boldsymbol{A}_{\mu },%
\boldsymbol{\psi },$ $\mathrm{\lambda })$ leads to following equations of
motion for the vector field multiplet $\boldsymbol{A}_{\mu }^{p}$ 
\begin{equation}
(\mathrm{E}_{\boldsymbol{A}}^{p})_{\mu }=0\text{ , \ }p=0,1,...,N-1
\label{w}
\end{equation}%
and the auxiliary field $\mathrm{\lambda }$%
\begin{equation}
\mathrm{C}(\boldsymbol{A})=\boldsymbol{A}_{\mu }^{p}\boldsymbol{A}^{p\mu }-%
\boldsymbol{n}^{2}\mathrm{M}^{2}=0  \label{constt}
\end{equation}%
which is the covariant length-preserving constraint for vector fields in the
non-Abelian symmetry case. Thus, there are a total of $N+1$ equations for $N$
4-vector fields $\boldsymbol{A}_{\mu }^{p}$, one of which is the constraint $%
\mathrm{C}(\boldsymbol{A})=0$ being preserved in time. This means, in
accordance with general arguments given above, that the equations of motion
for the vector fields $\boldsymbol{A}_{\mu }^{p}$ cannot be all independent.
As a result, the appropriate emergence equations, analogous to the equation (%
\ref{FF}) in an Abelian theory, inevitably occur%
\begin{equation}
\mathrm{F}^{p}(\mathrm{C}=0;\text{ \ }\mathrm{E}_{\boldsymbol{A}},\mathrm{E}%
_{\boldsymbol{\psi }})=0\text{, \ }p=0,1,...,N-1\text{ .}  \label{id000}
\end{equation}%
When being expressed as a linear combination of the basic mass dimension-4
terms, this equation leads to the identities between all field Eulerians
involved 
\begin{equation}
\partial _{\mu }(\mathrm{E}_{\boldsymbol{A}}^{p})^{\mu }=f^{pqr}\boldsymbol{A%
}_{\mu }^{q}(\mathrm{E}_{\boldsymbol{A}}^{r})^{\mu }+\mathrm{E}_{\boldsymbol{%
\psi }}(it^{p})\boldsymbol{\psi }+\overline{\boldsymbol{\psi }}(-it^{p})%
\mathrm{E}_{\overline{\boldsymbol{\psi }}}\text{ .}  \label{id111}
\end{equation}%
An identification of the coefficients of the Eulerians on the right-hand
side of the identities (\ref{id111}) with the structure constants $f^{pqr}$
and generators $t^{p}$ (\ref{22}) of the group $G$ is quite transparent.
This readily comes from the fact that the right-hand side of this identity
must transform in the same way as the left-hand side, which transforms as
the adjoint representation of the $G$ group. The terms and their
coefficients in the identity (\ref{id}) are typically chosen so as to
satisfy the Lee bracket operation to close the symmetry algebra once the
corresponding field transformations are identified\footnote{%
In order to avoid generating too many symmetry transformations, which would
only be consistent with a trivial Lagrangian (i.e. $\mathcal{L}=const$), we
typically require that the general symmetry transformations following from
the emergence identities like (\ref{id111}) have to constitute a group. This
means that they have to satisfy the Lie bracket operations to close the
symmetry algebra. This requirement puts strong restrictions on the form of
the emergence identities in all the emergent symmetry cases considered \cite%
{cfn2, c}.}. Note that, in contrast to the Abelian case, the term
proportional to the vector field multiplet $\boldsymbol{A}_{\mu }^{p}$
itself which corresponds to a self-interaction of non-Abelian vector fields,
also appears in the above identity. Again, Noether's second theorem \cite%
{noeth} can be applied directly to this identity in order to derive the
gauge invariance of the Lagrangian \textrm{L}$(\boldsymbol{A},\boldsymbol{%
\psi })$ in (\ref{33}). Indeed, with the constraint (\ref{constt}) implied,
the \textrm{L}$(\boldsymbol{A}_{\mu },\boldsymbol{\psi })$ tends to be
invariant under vector and fermion field local transformations having the
infinitesimal form 
\begin{equation}
\delta \boldsymbol{A}_{\mu }^{p}=\partial _{\mu }\boldsymbol{\omega }%
^{p}+f^{pqr}\boldsymbol{A}_{\mu }^{q}\boldsymbol{\omega }^{r},\text{ \ \ }%
\delta \boldsymbol{\psi }\text{\ }=(it^{p})\boldsymbol{\omega }^{p}%
\boldsymbol{\psi },\text{ \ \ }\delta \overline{\boldsymbol{\psi }}\text{\ }=%
\overline{\boldsymbol{\psi }}(-it^{p})\boldsymbol{\omega }^{p}.
\label{trans1}
\end{equation}%
For a theory with renormalizable coupling constants, this emergent gauge
symmetry leads to the conventional Yang-Mills type Lagrangian 
\begin{equation}
\mathrm{L}^{\mathfrak{em}}(\boldsymbol{A},\boldsymbol{\psi },\text{ }\mathrm{%
\lambda })=\mathrm{L}_{YM}(\boldsymbol{A},\boldsymbol{\psi })-\text{ }\frac{%
\mathrm{\lambda }}{2}(\boldsymbol{A}_{\mu }^{p}\boldsymbol{A}^{p\mu }-%
\boldsymbol{n}^{2}\mathrm{M}^{2})  \label{nab}
\end{equation}%
where we also include the corresponding Lagrange multiplier term. This term,
as was mentioned above, does not contribute to the vector field equation of
motion in the identity (\ref{id111}).

Now let us turn to the spontaneous symmetry violation which may be caused by
the nonlinear vector field constraint (\ref{constt}) determined by the
Lagrange multiplier term in (\ref{nab}). Although the Lagrangian \textrm{L}$%
^{\mathfrak{em}}(\boldsymbol{A},\boldsymbol{\psi },$ $\mathrm{\lambda })$
only has an $SO(1,3)\times G$ invariance, the constraint term in it
possesses the much higher accidental symmetry $SO(N,3N)$ according to the
dimension $N$ of the adjoint representation of $G$ to which the vector
fields $\boldsymbol{A}_{\mu }^{p}$ belong. This symmetry is indeed
spontaneously broken at a scale $\mathrm{M}$ 
\begin{equation}
\left\langle \boldsymbol{A}_{\mu }^{p}\right\rangle \text{ }=\boldsymbol{n}%
_{\mu }^{p}\mathrm{M}\text{ }  \label{5'}
\end{equation}%
with the vacuum direction determined now by the `unit' rectangular matrix $%
\boldsymbol{n}_{\mu }^{p}$ which describes simultaneously both of the
possible symmetry breaking cases, time-like 
\begin{equation}
SO(N,3N)\rightarrow SO(N-1,3N\mathbb{)}  \label{ss}
\end{equation}%
or space-like 
\begin{equation}
SO(N,3N)\rightarrow SO(N,3N-1)  \label{sss}
\end{equation}%
depending on the sign of $\boldsymbol{n}^{2}\equiv \boldsymbol{n}_{\mu }^{p}%
\boldsymbol{n}^{p\mu }=\pm 1$. In both cases the matrix $\boldsymbol{n}_{\mu
}^{p}$ has only one non-zero element, subject to the appropriate $SO(1,3)$
and (independently) $G$ rotations. They are, specifically, $\boldsymbol{n}%
_{0}^{0}$ or $\boldsymbol{n}_{3}^{0}$ provided that the VEV (\ref{5'}) is
developed along the $p=0$ direction in the internal space and along the $\mu
=0$ or $\mu =3$ direction, respectively, in the ordinary four-dimensional
spacetime.

As was argued in \cite{jej, cj}, side by side with one true vector Goldstone
boson corresponding to the spontaneous violation of the actual $%
SO(1,3)\otimes G$ symmetry of the Lagrangian \textrm{L}$^{\mathfrak{em}}(%
\boldsymbol{A}_{\mu },\boldsymbol{\psi },$ $\mathrm{\lambda })$, the $N-1$
pseudo-Goldstone vector bosons (PGB) related to the breakings (\ref{ss}, \ref%
{sss}) of the accidental symmetry $SO(N,3N)$ of the constraint (\ref{constt}%
) per se are also produced\footnote{%
Note that in total there appear $4N-1$ pseudo-Goldstone modes, complying
with the number of broken generators of $SO(N,3N)$. From these $4N-1$
pseudo-Goldstone modes, $3N$ modes correspond to the $N$ three-component
vector states as will be shown below, while the remaining $N-1$ modes are
scalar states which will be excluded from the theory.}. Remarkably, in
contrast to the familiar scalar PGB case \cite{GLA}, the vector PGBs remain
strictly massless being protected by the simultaneously generated
non-Abelian gauge invariance. Together with the above true vector Goldstone
boson, they also come into play properly completing the whole gauge
multiplet of the internal symmetry group $G$ taken.

Now we come again, as in the QED case, to the symmetry broken phase which in
accordance with the VEV equation (\ref{5'}) is defined as 
\begin{equation}
\text{\ \ }\boldsymbol{A}_{\mu }^{p}=\boldsymbol{a}_{\mu }^{p}+\boldsymbol{n}%
_{\mu }^{p}\mathrm{H}\text{ },\text{ \ }\boldsymbol{n}_{\mu }^{p}\boldsymbol{%
a}^{p\mu }\text{\ }=0\text{ \ \ \ }(\boldsymbol{a}^{2}\equiv \boldsymbol{a}%
_{\mu }^{p}\boldsymbol{a}^{p\mu })  \label{supp}
\end{equation}%
in which the new vector fields $\boldsymbol{a}_{\mu }^{p}$ appear as the
vector\ Goldstone modes. Indeed, this multiplet is orthogonal to the vacuum
direction given by the `unit' rectangular matrix $\boldsymbol{n}_{\mu }^{p}$%
, as is determined by the constraint (\ref{constt}) itself together with the
effective Higgs mode 
\begin{equation}
\mathrm{H}=\text{ }(\mathrm{M}^{2}-\boldsymbol{n}^{2}\boldsymbol{a}%
^{2})^{1/2}=\mathrm{M}-\frac{\boldsymbol{n}^{2}\boldsymbol{a}^{2}}{2\mathrm{M%
}}+\cdot \cdot \cdot  \label{H2}
\end{equation}%
Note that, apart from the pure vector fields, general zero modes $%
\boldsymbol{a}_{\mu }^{p}$ contain $N-1$ scalar modes, $\boldsymbol{a}%
_{0}^{p^{\prime }}$ or $\boldsymbol{a}_{3}^{p^{\prime }}$ ($p^{\prime
}=1,...,N-1$), for the time-like ($\boldsymbol{n}_{\mu }^{p}=n_{0}^{0}g_{\mu
0}\delta ^{p0}$) or space-like ($\boldsymbol{n}_{\mu }^{p}=n_{3}^{0}g_{\mu
3}\delta ^{p0}$) SLIV, respectively. They can be eliminated from the theory,
if one imposes appropriate supplementary conditions on the $N-1$ fields $%
\boldsymbol{a}_{\mu }^{p}$ which are still free of constraints. Using their
overall orthogonality (\ref{supp}) to the physical vacuum direction $%
\boldsymbol{n}_{\mu }^{p}$, one can formulate these supplementary conditions
in terms of a general axial gauge for the entire $\boldsymbol{a}_{\mu }^{p}$
multiplet 
\begin{equation}
n\cdot \boldsymbol{a}^{p}\equiv n_{\mu }\boldsymbol{a}^{p\mu }=0,\text{ \ }%
p=0,1,...,N-1.  \label{supp'}
\end{equation}%
Here $n_{\mu }$ is the unit Lorentz vector being analogous to the vector
introduced in the Abelian case, which is now oriented in Minkowski spacetime
so as to be "parallel" to the vacuum unit $\boldsymbol{n}_{\mu }^{p}$
matrix. This matrix can be taken hereafter in the "two-vector" form 
\begin{equation}
\boldsymbol{n}_{\mu }^{p}=n_{\mu }\boldsymbol{\epsilon }^{p}\text{ },\text{ }%
\boldsymbol{\epsilon }^{p}\boldsymbol{\epsilon }^{p}=1
\end{equation}%
where $\boldsymbol{\epsilon }^{p}$ is unit $G$ group vector belonging to its
adjoint representation. As a result, in addition to an elementary Higgs mode
excluded earlier by the above orthogonality condition (\ref{supp}), all the
other scalar fields are also eliminated. Consequently only the pure vector
fields, $\boldsymbol{a}_{i}^{p}$ ($i=1,2,3$ ) or $\boldsymbol{a}_{\mu
^{\prime }}^{p}$ ($\mu ^{\prime }=0,1,2$), for the time-like or space-like
SLIV respectively, are left in the theory. Clearly, the components $%
\boldsymbol{a}_{i}^{p=0}$ and $\boldsymbol{a}_{\mu ^{\prime }}^{p=0}$
correspond to the true Goldstone bosons in these cases, while all the other
(for $p=1,...,N-1$) are vector PGBs.

Substituting the parameterization (\ref{supp}, \ref{H2}) into the Lagrangian
(\ref{nab}), one is led to the non-Abelian gauge theory in terms of the pure
emergent modes $\boldsymbol{a}_{\mu }^{p}$. However, as in the above Abelian
case, one should first use the local invariance of the Yang-Mills Lagrangian 
$\mathrm{L}_{YM}(\boldsymbol{A},\boldsymbol{\psi })$ in the emergent theory (%
\ref{nab}) to gauge away the apparently large but fictitious Lorentz
violating terms which appear in the symmetry broken phase (\ref{supp}). As
one can readily see, they stem from the effective Higgs field $\mathrm{H}$
expansion (\ref{H2}) when it is applied to some vector-vector and
vector-fermion field interaction couplings in the Lagrangian $\mathrm{L}%
_{YM} $ in (\ref{nab}). To exclude them we can make the appropriate
transformations (similar to the transformation (\ref{red}) taken above in
the Abelian case) of the fermion ($\boldsymbol{\psi }$) and new vector ($%
\boldsymbol{a}_{\mu }\equiv \boldsymbol{a}_{\mu }^{p}t^{p}$) field
multiplets: 
\begin{equation}
\boldsymbol{\psi }\rightarrow U(\omega )\boldsymbol{\psi }\text{ },\text{ \
\ }\boldsymbol{a}_{\mu }\rightarrow U(\omega )\boldsymbol{a}_{\mu }U(\omega
)^{\dagger },\text{ \ }U(\omega )=e^{ig\mathrm{M}(x\cdot \boldsymbol{n})}%
\text{.}  \label{red1}
\end{equation}%
Since the phase of the transformation phase $\omega $ is linear in the
spacetime coordinate the following equalities are evidently satisfied: 
\begin{equation}
\partial _{\mu }U(\omega )=ig\mathrm{M}\boldsymbol{n}_{\mu }U(\omega )=ig%
\mathrm{M}U(\omega )\boldsymbol{n_{\mu }},\text{ \ \ }\boldsymbol{n}_{\mu
}\equiv \boldsymbol{n}_{\mu }^{p}t^{p}.  \label{r2}
\end{equation}%
One can readily confirm now that the above-mentioned fictitious Lorentz
violating terms in Lagrangian (\ref{nab}) are thereby cancelled with the
analogous terms stemming from kinetic terms of vector and fermion fields.
So, the final Lagrangian for the emergent Yang-Mills theory in the symmetry
broken phase takes the form 
\begin{eqnarray}
\mathrm{L}^{\mathfrak{em}}(\boldsymbol{a},\boldsymbol{\psi }) &=&-\frac{1}{4}%
[\boldsymbol{f}_{\mu \nu }^{p}+\overline{\boldsymbol{f}}_{\mu \nu }^{p}(%
\mathrm{H}-\mathrm{M})][\boldsymbol{f}^{p\mu \nu }+\overline{\boldsymbol{f}}%
^{p\mu \nu }(\mathrm{H}-\mathrm{M})]-\frac{1}{2}\delta (n^{\mu }\boldsymbol{a%
}_{\mu }^{p})^{2}  \notag \\
&&+\overline{\boldsymbol{\psi }}(i\gamma \partial -m)\boldsymbol{\psi }+g[%
\boldsymbol{a}_{\mu }^{p}+(\mathrm{H}-\mathrm{M})\boldsymbol{n}_{\mu }^{p}]%
\overline{\boldsymbol{\psi }}(\gamma ^{\mu }t^{p})\boldsymbol{\psi }\text{ .}
\label{nab2}
\end{eqnarray}%
Here the stress-tensor $\boldsymbol{f}_{\mu \nu }^{p}$ is, as usual, 
\begin{equation}
\boldsymbol{f}_{\mu \nu }^{p}\boldsymbol{~=~}\partial _{\mu }\boldsymbol{a}%
_{\nu }^{p}-\partial _{\nu }\boldsymbol{a}_{\mu }^{p}+gf^{pqr}\boldsymbol{a}%
_{\mu }^{q}\boldsymbol{a}_{\nu }^{r},  \label{dmm}
\end{equation}%
while $\overline{\boldsymbol{f}}_{\mu \nu }^{p}$ stands for the new SLIV
oriented tensor 
\begin{equation}
\overline{\boldsymbol{f}}_{\mu \nu }^{p}\equiv \boldsymbol{n}_{\nu
}^{p}\partial _{\mu }-\boldsymbol{n}_{\mu }^{p}\partial _{\nu }+gf^{pqr}(%
\boldsymbol{a}_{\mu }^{q}\boldsymbol{n}_{\nu }^{r}-\boldsymbol{a}_{\nu }^{q}%
\boldsymbol{n}_{\mu }^{r}).  \label{dm}
\end{equation}%
acting on the effective Higgs field expansion terms in (\ref{nab2}). We also
included the overall gauge fixing term for the entire $\boldsymbol{a}_{\mu
}^{p}$ multiplet to remove all scalar modes from the theory and retained the
original notations for the fermion and vector fields after the
transformations (\ref{red1}) were carried out. One can easily see that these
transformations actually amounts to the replacement of the effective Higgs
field $\mathrm{H}$ by the combination $\mathrm{H}-\mathrm{M}$ in the
emergent Lagrangian (\ref{nab2}). As a result, all the large Lorentz
breaking terms proportional to the scale $\mathrm{M}$ appear cancelled%
\footnote{%
Let us note that the constant background value of the gauge vector field can
always be cancelled out depending no whether the Higgs field is effective or
elementary. Indeed, the vacuum shift of a vector field in an Abelian case is
a gauge transformation in itself with a gauge function linear in coordinate,
as can be readily seen in (\ref{ab}). In non-Abelian case, the vacuum shift
of a vector field multiplet $\boldsymbol{A}_{\mu }$ given in (\ref{supp}, %
\ref{H2}) being accompanied by a subsequent "rotation" (\ref{red1}, \ref{r2}%
) of an emergent multiplet $\boldsymbol{a}_{\mu }$ can be written entirely as%
\begin{equation*}
\boldsymbol{A}_{\mu }=U\boldsymbol{a}_{\mu }U^{-1}-\frac{i}{g}(\partial
_{\mu }U)U^{-1}+\boldsymbol{n}_{\mu }(H-M).
\end{equation*}%
with the transformation phase linear in the coordinate. Now one can clearly
see that the first two terms present the pure gauge transfomation of vector
field multiplet $\boldsymbol{a}_{\mu }$ and, thereby, only the third term
can not be gauged away. That is why the emergent Lagrangian (\ref{nab2})
contains just the combintion $H-M$. Due to this fact the Lagrangian $\mathrm{%
L}^{\mathfrak{em}}$ goes to a conventional Yang-Mills theory in the limit $%
\boldsymbol{M\rightarrow \infty }$. Otherwise, it would become ifinite in
this limit.}. Expanding the effective Higgs field $\mathrm{H}$ (\ref{H2}) in
powers of $\boldsymbol{a}^{2}/\mathrm{M}^{2}$ in the Lagrangian (\ref{nab2}%
), one comes to highly nonlinear theory in terms of the zero vector modes $%
\boldsymbol{a}_{\mu }^{p}$ which contains many the properly suppressed
Lorentz violating couplings. However, as in the Abelian symmetry case, they
do not lead to physical Lorentz violation effects which turn out to be
strictly cancelled among themselves \cite{jej, cj}.

All the above allows one to conclude that the Yang-Mills theory can
naturally be interpreted as emergent gauge theory caused by the generic
length-preserving constraint put on the vector field multiplet in some
prototype theory possessing only global non-Abelian internal symmetry. Such
a constraint could in principle lead to the spontaneous Lorentz violation
which, however, appears unobservable due to the gauge invariance (\ref%
{trans1}) emerged, thus giving one more example of an inactive SLIV. In this
connection, the emergence conjecture itself could be reformulated as a
principle \cite{cfn} of a generic non-observability of the spontaneous
Lorentz violation being caused by some constant background value of the
vector field (or vector field multiplet). Presumably, this principle may
provide the origin of all gauge internal symmetries observed whether they
are exact as in QED and quantum chromodynamics or spontaneously broken as in
the electroweak theory and grand unified models.

\section{Towards an emergent \textbf{Poincar\'{e}} gravity}

We have mentioned above that PGG looks in essence as a gauge field theory in
flat Minkowski space which successfully mimics curved space geometry when
making the transition to the base world space with general affine
connections and metric expressed, respectively, in spin-connections and
tetrads. Conventionally, one has in PGG the world space (WS) symmetry\ $%
ISO(1,3)_{WS}$, which includes translations and the orbital part of Lorentz
transformations, and a local frame \ (LF)\ Lorentz symmetry $SO(1,3)_{LF}$,
which only includes the spin part of Lorentz transformations acting on
representation indices. {}Remarkably, this duality is in an automatic
accordance with the Einstein equivalence principle which, therefore, need
not to be specially postulated in PGG\ as is in the standard GR. On the
other hand, this suggests some clear analogy between the local frame
symmetry space in PGG and internal charge space in conventional quantum
field theories. Such a unique property of PGG allows us to proceed in
precisely the same way as before in the Yang-Mills theory case that
eventually leads us to the emergent PGG theory. In this connection, we begin
with the entirely global spacetime symmetries, both $ISO(1,3)_{WS}$ and $%
SO(1,3)_{LF}$, and our starting objects are the two vector field multiplets
which are 4-vectors of $ISO(1,3)_{WS}$ and belong, respectively, to the
adjoint ($A_{\mu }^{ij}$) and vector ($e_{\mu }^{i}$) representations of the
Lorentz group $SO(1,3)_{LF}$ (antisymmetry in the indices $ij$ in $A_{\mu
}^{ij}$ is imposed). In what follows, we will refer to these prototype
fields as the spin connections and tetrads. As we will see, they are really
turned out to be those when an emergence procedure given above in Section II
B is applied to them. As a result, the local frame Lorentz symmetry $%
SO(1,3)_{LF}$ and translation subgroup in $ISO(1,3)_{WS}$ appear gauged,
while the orbital Lorentz transformations are actually absorbed by the
latter. So, eventually one has the local translations and local $SO(1,3)_{LF}
$ transformations being gauged by the emergent tetrad and spin-connection
fields, respectively. The gauge tetrad field $e_{\mu }^{i}$ set the local
coordinate frame in the emergent PGG, while the gauge spin-connection fields 
$A_{\mu }^{ij}$ are in fact connections with respect to the tetrad frame (we
retain the starting notations for them). Again, due to gauge invariance
emerged the physical Lorentz (and translation invariance) remains in the
final theory. This means that the covariant length-preserving constraints
only lead to an inactive SLIV case even if they are applied to vector fields
related to spacetime symmetries.

\subsection{Constrained tetrad and spin-connection fields}

First of all, as we could learn above from the emergent gauge theories, the
tetrad and spin-connection fields have to be properly constrained to induce
an approprite emergence process. The essential point is, however, that the
tetrad field is generically constrained by definition (\ref{t}). To see
clearer what does this constraint mean, let first notice that whereas the
spin-connection field $A_{\mu \text{ \ \ }}^{ij}$has a canonical vector
field mass dimension, the tetrad field $e_{\mu }^{i}$ appears to have zero
mass dimension. Treating it as all other boson fields having a canonical
dimension of mass we introduce some fundamental mass scale in the above
definition of tetrad fields $e_{\mu }^{i}$ (\ref{t}) changing their
orthogonality equations to%
\begin{equation}
e_{\mu }^{i}e_{i}^{\nu }\,=\delta _{\mu }^{\nu }M_{e}^{2}\text{ , \ }~e_{\mu
}^{i}e_{j}^{\mu }\,=\delta _{j}^{i}\,M_{e}^{2}\text{ , \ \ }e_{\mu
}^{i}e_{i}^{\mu }\,=\mathrm{n}^{2}M_{e}^{2}\text{ ,\ }  \label{t1}
\end{equation}%
where the first two conditions could be considered as those which define the
inverse tetrads $e_{i}^{\mu }$, whereas the third one is their length-fixing
constraint. Here $\mathrm{n}^{2}$ stands for 
\begin{equation}
\mathrm{n}^{2}\equiv \delta _{\mu }^{\nu }\delta _{\nu }^{\mu }=\delta
_{j}^{i}\delta _{i}^{j}=\delta _{\mu }^{i}\delta _{i}^{\mu }=4\text{ .}
\label{t11}
\end{equation}%
Since a general metric tensor $g_{\mu \nu }$ is generally assumed to be
dimensionless one also has%
\begin{equation}
g_{\mu \nu }=\frac{1}{M_{e}^{2}}\eta _{ij}e_{\mu }^{i}e_{\nu }^{j}\text{ .}
\label{t2}
\end{equation}%
rather than (\ref{metric}). We can readily see that the last constraint in (%
\ref{t1}) is indeed similar to the constraints we have above for
conventional vector fields (\ref{const}, \ref{constt}). This constraint
actually means that PGG is a spontaneously broken theory that manifests
itself at some input mass scale $M_{e}$ which could be in principle
associated with the Plank mass $M_{P}$. Generally, this violation may
concern both the world spacetime symmetry\ and the local frame Lorentz
symmetry being developed along some particular directions, just as it took
place in the non-Abelian vector field case considered above. However, one
can choose this violation in a way that the vacuum of the PGG theory is flat
Minkowski space rather than breaks Lorentz invariance. This lead, as we will
see shortly, to spontaneous violation of some accidental global symmetry and
generation of the Goldstone vector bosons properly gauging translation in
the world space.

The similar length-fixing constraint is proposed to be put on the
spin-connection fields $A_{\mu }^{ij}$%
\begin{equation}
A_{\mu }^{ij}A_{ij}^{\mu }=n^{2}M_{A}^{2}\text{\ , \ \ }n^{2}\equiv n_{\mu
}^{ij}n_{ij}^{\mu }=\pm 1  \label{a1}
\end{equation}%
being analogous to the constraints (\ref{const}, \ref{constt}) for ordinary
vector fields (here $n_{\mu }^{ij}$ stands now for some properly-oriented
`unit' rectangular matrix, and also antisymmetry in the $ij$ indices is
imposed). The constraint\ (\ref{a1}) actually means that we also have a
spontaneous Lorentz violation in PGG that appears at some high mass scale $%
M_{A}$ which could be in principle close to the Plank mass $M_{P}$ as well.
This will cause, as we confirm later, the generation of Goldstone vector
bosons gauging Lorentz symmetry in the local frame, while the physical
Lorentz invariance is left intact. Remarkably, we have here the very special
case of an inactive SLIV which induces a local Lorentz invariance.

\subsection{From global to local symmetries}

We have already emphasized above that there is a generic analogy between the
local frame symmetry space in PGG and internal charge space in conventional
quantum field theories. As a result, the vector fields of spin-connections $%
A_{\mu }^{ij}(x)$ gauging the local Lorentz group $SO(1,3)_{LF}$ look like
as gauge bosons appearing in the Standard Model. As to tetrads $e^{i}{_{\mu
}(x)}$, though they are gauge fields of the coordinate-dependent
translations in the world spacetime, they transform like as ordinary matter
fields w.r.t. the local Lorentz frame. As such, they belong to the vector
multiplet of $SO(1,3)_{LF}$ rather than its adjoint representation as the
spin-connection fields $A_{\mu }^{ij}(x)$.

All that looks very similar to the situation we had above in the Yang-Mills
theory case, and actually represents some precondition for PGG to be also
emerged from pure global symmetries once the vector field constraints (\ref%
{t1}) and (\ref{a1}) come into play. Accordingly, we start with some
prototype theory possessing only global symmetries $ISO(1,3)_{WS}$ and $%
SO(1,3)_{LF}$ operating in the two flat Minkowski spaces with constant
metrics $\eta _{\mu \nu }$ and $\eta _{ij}$, respectively. This yet
arbitrary theory contains some prototype vector fields having form of
spin-connections $A_{\mu }^{ij}(x)$ and tetrads $e_{\mu }^{i}{(x)}$ and may
also contain some matter fields (say, fermions $\psi $). The theory have in
general all possible interactions between all vector and matter fields
involved. The corresponding Lagrangian $\mathcal{L}^{tot}$ is supposed to
also include the standard Lagrange multiplier terms with the field functions 
$\lambda _{A}(x)$ and $\lambda _{e}(x)$ 
\begin{equation}
\mathcal{L}^{tot}(e,A{,}\psi ;\lambda _{e},\lambda _{A})=\mathcal{L}(e,A{,}%
\psi )-\text{ }\frac{\lambda _{A}}{2}(A_{\mu }^{ij}A_{ij}^{\mu
}-n^{2}M_{A}^{2})-\frac{\lambda _{e}}{2}(e_{\mu }^{i}\,e_{i}^{\mu }-\mathrm{n%
}^{2}M_{e}^{2})\text{.}  \label{l}
\end{equation}%
As in the above QED and Yang-Mills theory cases (and for the same reason,
see also footnote$^{3}$), these terms do not contribute to the vector field
equations of motion. The variations under $\lambda _{A}(x)$ and $\lambda
_{e}(x)$ result, accordingly, in the covariant length-preserving constraints
for the spin-connection and tetrad fields 
\begin{equation}
\mathcal{C}_{A}=A_{\mu }^{ij}A_{ij}^{\mu }-n^{2}M_{A}^{2}=0\text{\ , \ }%
\mathcal{C}_{e}=e_{\mu }^{i}e_{i}^{\mu }\,-\mathrm{n}^{2}M_{e}^{2}=0
\label{cc}
\end{equation}%
in the PGG theory. Therefore, we again face the question of consistency of
these extra constraint equations with the equations of motion for the vector
fields of tetrads $e_{\mu }^{i}$ and spin-connections $A_{\mu }^{ij}$%
\begin{equation}
(\mathcal{E}_{A}^{ij})_{\mu }=0\text{ , \ }(\mathcal{E}_{e}^{j})_{\mu }=0%
\text{ \ }(i,j=0,1,2,3;\text{ }\mu =0,1,2,3)\text{ .}  \label{q}
\end{equation}%
For an arbitrary Lagrangian $\mathcal{L}(e,A{,}\psi ),$ the time development
of the fields would not preserve in general the constraints (\ref{cc}). So,
the parameters in the Lagrangian $\mathcal{L}$ must be chosen so as to give
a relationship between the Eulerians for all the fields involved. The need
to preserve the constraints $\mathcal{C}_{A}=0$ and $\mathcal{C}_{e}=0$ with
time implies, as in the emergent Yang-Mills theory case, that the equations
of motion for the vector fields of spin-connections $A_{\mu }^{ij}$ and
tetrads $e_{\mu }^{i}$, respectively, cannot be all independent. As a
result, the special emergence equations for spin-connection fields 
\begin{equation}
\mathcal{F}^{ij}(\mathcal{C}_{A}=0;\text{ }\mathcal{E}_{A},\mathcal{E}_{e},%
\mathcal{E}_{\psi },...)=0\text{ }(i,j=0,1,2,3)\text{ }  \label{f}
\end{equation}%
and tetrad fields%
\begin{equation}
\mathcal{F}_{\mu }(\mathcal{C}_{e}=0;\text{ }\mathcal{E}_{e},\mathcal{E}_{A},%
\mathcal{E}_{\psi },...)=0\text{ \ \ \ }(\mu =0,1,2,3),  \label{fmu}
\end{equation}%
necessarily appear. Remind one more time that an antisymmetry in the Lorentz
indices ($i,j,k,...$) is everywhere imposed.

Let us consider first the emergence equations (\ref{f}). Again, when being
expressed as a linear combination of the basic mass dimension-4 terms, this
equation leads to the identities between all field Eulerians involved 
\begin{equation}
\partial ^{\mu }(\mathcal{E}_{A})_{\mu }^{ij}=c_{\text{ }[kl][mn]\text{\ }%
}^{[ij]}A_{\mu \text{ \ \ }}^{kl}(\mathcal{E}_{A})^{\mu ,mn}+e_{\mu }^{[i}(%
\mathcal{E}_{e})^{j],\mu }+\mathcal{E}_{\psi }S^{ij}\psi +\overline{\psi }%
S^{ij}\mathcal{E}_{\overline{\psi }}  \label{id}
\end{equation}%
which are precisely analogous to those we had above in (\ref{id111}) for the
Yang-Mills theory. An appropriate identification of the Eulerian terms on
the right-hand side of the identity (\ref{id}) with the structure constants $%
c_{\text{ }[kl][mn]\text{\ }}^{[ij]}$and the fermion representation matrices 
$S^{ij}$ of the Lorentz symmetry group $SO(1,3)_{LF}$ is indeed quite clear.
The point is that the right-hand side of this identity must transform in the
same way as its left-hand side, which transforms as the adjoint
representation of $SO(1,3)_{LF}$. As to their coefficients and other
possible terms in the identity (\ref{id}), there remained only terms which
satisfy the Lee bracket operation to close the symmetry algebra once the
corresponding field transformations are identified (in this connection, see
our comment to the Yang-Mills theory case in the footnote$^{6}$).

As to the basic identities following from the emergence equations for tetrad
fields (\ref{fmu}), the non-trivial lowest mass dimension terms constructed
from the Eulerians for this case will necessarily include the translation
operator expression $T_{\mu }=-\partial _{\mu }$ for all the fields
involved. Consequently they take the following form 
\begin{equation}
e_{\nu }^{i}\,\partial _{\mu }(\mathcal{E}_{e})\,_{i}^{\nu }+A_{\nu
}^{ij}\partial _{\mu }(\mathcal{E}_{A})_{ij}^{\nu }+(\partial _{\mu }%
\mathcal{E}_{\psi })\psi +\overline{\psi }(\partial _{\mu }\mathcal{E}_{%
\overline{\psi }})=0  \label{idd}
\end{equation}%
which consist of all the terms having mass dimension 5.

Now again, Noether's second theorem \cite{noeth} can be applied directly to
the above identities (\ref{id}) and (\ref{idd}) in order to derive the gauge
invariance of the Lagrangian $\mathcal{L}(e,A{,}\psi )$ in (\ref{l}).
Indeed, with the constraint (\ref{cc}) implied, this Lagrangian tends to be
invariant under local transformations of the spin-connection, tetrad and
matter fields of the type%
\begin{eqnarray}
\delta A_{\mu }^{ij} &=&\varepsilon _{k}^{i}\,A_{\mu }^{kj}\,+\varepsilon
_{k}^{j}\,A_{\mu }^{ik}\,-\partial _{\mu }\xi ^{\nu }A_{\nu
\;}^{ij}-\partial _{\mu }\varepsilon ^{ij},  \notag \\
\delta e_{i}^{\mu }\, &=&e_{i}^{\nu }\,\partial _{\nu }\xi ^{\mu
}-e_{k}^{\mu }\,\varepsilon _{i}^{k}\,\text{ , }\delta \psi =\frac{1}{2}%
\varepsilon ^{ij}S_{ij}\psi \text{ .}  \label{A}
\end{eqnarray}%
Note that the first two terms in $\delta A_{\mu }^{ij}$ correspond to local
Lorentz rotations of the spin-connection fields $A_{\mu }^{ij}$ with
parameters $\varepsilon ^{ij}(x)$, while the third term is due to the local
translations conditioned by the parameters $\xi ^{\mu }(x)$. The last terms
in $\delta A_{\mu }^{ij}$ means that the spin-connection fields $A_{\mu
}^{ij}$ gauge just the local Lorentz rotations. The tetrad field in $\delta
e_{i}^{\mu }$ is Lorentz-rotated (in the local Lorentz frame) and,
simultaneously, subject to the coordinate-dependent translations (in the
world spacetime). And finally, the transformation of the fermion field $\psi 
$ in (\ref{A}) is, as usual, determined by the fermion representation
matrices $S^{ij}$ which are simply given by the commutators of the $\gamma $%
-matrices 
\begin{equation}
S_{ij}=\frac{1}{4}\left[ \gamma _{i}\text{, }\gamma _{j}\right] \text{ \ }%
\equiv \gamma _{ij}/2\text{ }.  \label{12}
\end{equation}%
The local transformations (\ref{A}) shows that the somewhat arbitrarily
introduced prototype vector fields $A_{\mu }^{ij}$ and $e_{\mu }^{i}$ are
really turned out to be the PGG spin-connection and tetrad fields once they
satisfy the length-preserving constraints (\ref{cc}). Moreover, the induced
gauge symmetry (\ref{A}) unavoidably leads to the emergent PGG Lagrangian%
\begin{equation}
\mathcal{L}^{\mathfrak{em}}(e,A{,}\psi ;\lambda _{e},\lambda _{A})=\mathcal{L%
}^{\mathfrak{em}}(e,A{,}\psi )_{PGG}-\text{ }\frac{\lambda _{A}}{2}(A_{\mu
}^{ij}A_{ij}^{\mu }-n^{2}M_{A}^{2})-\frac{\lambda _{e}}{2}(e_{\mu
}^{i}\,e_{i}^{\mu }-\mathrm{n}^{2}M_{e}^{2})  \label{l'}
\end{equation}%
where $\mathcal{L}^{\mathfrak{em}}(e,A{,}\psi )_{PGG}$ is solely constructed
from the covariant curvature and torsion tensors%
\begin{equation}
R_{\mu \nu }^{ij}\,=\partial _{\lbrack \nu }A_{\mu ]}^{ij}+\eta _{kl}A_{%
\text{ }[\nu }^{ik}A_{\text{ \ \ }\mu ]}^{lj}\text{ , }T_{\mu \nu
}^{i}=\partial _{\lbrack \nu }e_{\mu ]}^{i}+\eta _{kl}A_{[\nu }^{ik}\,e_{\mu
]}^{l}  \label{cov}
\end{equation}%
and a covariant derivative for the fermion field 
\begin{equation}
\bar{\psi}\gamma ^{i}\overleftrightarrow{D}_{\mu }\psi =\bar{\psi}\gamma
^{i}(\partial _{\mu }\psi )-(\partial _{\mu }\bar{\psi})\gamma ^{i}\psi 
\text{ }+\frac{1}{4}A_{\mu }^{ab}\bar{\psi}\{\gamma ^{i},\gamma _{ab}\}\psi
\label{de}
\end{equation}%
We also included the corresponding Lagrange multiplier terms which, as was
mentioned above, do not contribute to the physical field equations of
motion. Now, for a theory with the lowest dimension coupling constants,
containing at most the quadratic terms in the curvature and torsion one has%
\begin{equation}
\mathcal{L}^{\mathfrak{em}}(e,A{,}\psi )_{PGG}=\mathcal{L}^{(1)}(e,A{,}\psi
)+\mathcal{L}^{(2)}(e,A{,}\psi )  \label{larg}
\end{equation}%
where the first term correspond to the minimal Einstein-Cartan theory being
linear in the curvature%
\begin{equation}
\mathcal{L}^{(1)}(e,A{,}\psi )=\frac{e}{2\kappa }\frac{e_{i}^{\mu
}e_{j}^{\nu }}{M_{e}^{2}}R_{\mu \nu }^{ij}+e\frac{e_{i}{}^{\mu }}{2M_{e}}%
\bar{\psi}\gamma ^{i}\overleftrightarrow{D}_{\mu }\psi  \label{ll}
\end{equation}%
(where $\kappa $ stands for the modified Newtonian constant $8\pi G$), while
in the second term $\mathcal{L}^{(2)}$ all eight possible quadratic terms 
\cite{nev1, nev} are generally collected.

\subsection{Broken symmetry phase: zero spin-connection modes}

We have found above that the presence of the spin-connection and tetrad
field constraints (\ref{cc}) in the theory unambiguously converts the global
symmetry $ISO(1,3)_{WS}\times SO(1,3)_{LF}$ we started into the local Poincar%
\'{e} symmetry $T(1,3)_{WS}\times SO(1,3)_{LF}$ that leads to the
conventional PGG theory\footnote{%
The orbital part of Lorentz symmetry transformations in the starting $%
ISO(1,3)_{WS}$ group is in fact absorbed by local translations, as mentioned
above.}. The point is, however, that these constraints mean at the same time
that this global symmetry is spontaneously broken thus inducing the
Goldstone spin-connection and tetrad field modes in which the PGG theory has
to be eventually expressed.

To see it in more detail, let us consider first the spin-connection fields.
Note above all, whereas the emergent PGG Lagrangian $\mathcal{L}_{PGG}^{%
\mathfrak{em}}$ in (\ref{l'}) possesses the local Poincar\'{e} symmetry $%
T(1,3)_{WS}\times SO(1,3)_{LF}$, the accidental global symmetry of the
length-fixing spin-connection constraint (\ref{a1}) appears much higher, $%
ISO(6,18)_{WS}$ \footnote{%
This symmetry being treated as the world space symmetry is determined by a
proper number of the spacetime directions related to the (local frame)
Lorentz group representations of the vector fields involved (just like as in
the Yang-Mills theory case considered above such a total symmetry was
determined by a number of the internal space directions related to an
adjoint representation of the vector field multiplet involed). In this way,
the length-fixing constraint for spin-connection fields (\ref{a1}) possesses
the global symmetry $ISO(6,18)_{WS}$, whereas a similar constraint for
tetrad fields (\ref{t1}) the lower global symmetry $ISO(4,12)_{WS}$, as is
claimed below.}. This symmetry is indeed spontaneously broken at a scale $%
M_{A}$ 
\begin{equation}
\left\langle A_{\mu }^{ij}\right\rangle \text{ }=n_{\mu }^{ij}M_{A}\text{ }
\label{v1}
\end{equation}%
with the vacuum direction determined now by the `unit' rectangular matrix $%
n_{\mu }^{ij}$ which describes simultaneously both of the SLIV cases,
time-like 
\begin{equation}
ISO(6,18)\rightarrow ISO(5,18)  \label{i1}
\end{equation}%
or space-like 
\begin{equation}
ISO(6,18)\rightarrow ISO(6,17)  \label{i2}
\end{equation}%
depending on the sign of $n^{2}=\pm 1$. In both cases the matrix $n_{\mu
}^{ij}$ has only one non-zero element, subject to the appropriate $%
ISO(1,3)_{WS}$ and (independently) $SO(1,3)_{LF}$ transformations. They are,
specifically, $n_{0}^{\left\langle ij\right\rangle }$ or $%
n_{3}^{\left\langle ij\right\rangle }$ provided that the VEV (\ref{v1}) is
developed along the $\left\langle ij\right\rangle $ direction in the local
Lorentz frame and along the $\mu =0$ or $\mu =3$ direction, respectively, in
the world spacetime.

As was argued in the above non-Abelian vector field case, side by side with
one true vector Goldstone boson corresponding to spontaneous violation of an
actual $ISO(1,3)_{WS}\times SO(1,3)_{LF}$ symmetry of the PGG\ Lagrangian,
the five pseudo-Goldstone vector bosons related to the breakings (\ref{i1}, %
\ref{i2}) of the accidental symmetry $ISO(6,18)$ of the constraint (\ref{a1}%
) per se are also produced\footnote{%
Note that in total there appear the $23$ pseudo-Goldstone modes, complying
with the number of broken generators of $SO(6,18)$. From these $23$
pseudo-Goldstone modes, $18$ modes correspond to the six three-component
vector states, as will be shown below, while the remaining $5$ modes are
scalar states which will be excluded from the theory.}. Remarkably, the
vector PGBs remain strictly massless being protected by the simultaneously
generated Lorentz gauge invariance. Together with the above true vector
Goldstone boson, they also come into play thus properly completing the
entire adjoint gauge multiplet of spin-connection fields of the local
Lorentz symmetry group $SO(1,3)_{LF}$.

Due to the constraint (\ref{a1}), which virtually appears as a single
condition put on the spin-connection field multiplet $A_{\mu }^{ij}$, one
can identify the pure Goldstone field modes $\mathcal{A}_{\mu }^{ij}$ using
the parametrization 
\begin{equation}
\text{\ \ }A_{\mu }^{ij}=\mathcal{A}_{\mu }^{ij}+n_{\mu }^{ij}\sqrt{%
M_{A}^{2}-n^{2}\mathcal{A}^{2}}\text{ },\text{ \ }n_{\mu }^{ij}\mathcal{A}%
_{ij}^{\mu }\text{\ }=0\text{ \ \ \ }(\mathcal{A}^{2}\equiv \mathcal{A}_{\mu
}^{ij}\mathcal{A}_{ij}^{\mu }).  \label{a2}
\end{equation}%
and an effective \textquotedblleft Higgs" mode 
\begin{equation}
\mathcal{H}=\sqrt{M_{A}^{2}-n^{2}\mathcal{A}^{2}}=M_{A}-\frac{n^{2}\mathcal{A%
}^{2}}{2M_{A}}+\cdot \cdot \cdot  \label{hm}
\end{equation}%
determined by the constraint itself. Note that, apart from the pure vector
fields, the general zero modes $\mathcal{A}_{\mu }^{ij}$ contain the five
scalar modes, $\mathcal{A}_{0}^{ij}$ or $\mathcal{A}_{3}^{ij}$, for the
time-like ($n_{\mu }^{ij}=n_{0}^{\left\langle ij\right\rangle }g_{\mu
0}\delta ^{(ij)\left\langle ij\right\rangle }$) or space-like ($n_{\mu
}^{ij}=n_{3}^{\left\langle ij\right\rangle }g_{\mu 3}\delta
^{(ij)\left\langle ij\right\rangle }$) SLIV, respectively. They can be
eliminated from the theory, if one imposes appropriate supplementary
conditions on the five fields $\mathcal{A}_{\mu }^{ij}$ which are still free
of constraints. Using their overall orthogonality (\ref{a2}) to the physical
vacuum direction $n_{\mu }^{ij}$, one can formulate these supplementary
conditions in terms of a general axial gauge for the entire $\mathcal{A}%
_{\mu }^{ij}$ multiplet 
\begin{equation}
n^{\mu }\mathcal{A}_{\mu }^{ij}=0.  \label{a3}
\end{equation}%
Here $n_{\mu }$ is the unit world spacetime vector which is oriented so as
to be "parallel" to the vacuum unit $n_{\mu }^{ij}$ matrix. This matrix can
be taken hereafter in the "two-vector" form 
\begin{equation}
n_{\mu }^{ij}=n_{\mu }\mathfrak{\epsilon }^{ij}\text{ \ \ (}n_{\mu }n^{\mu
}=1,\text{ }\mathfrak{\epsilon }^{ij}\mathfrak{\epsilon }_{ij}=1\text{)}
\label{vec}
\end{equation}%
where $\mathfrak{\epsilon }^{ij}$ is the unit Lorentz group tensor belonging
to its adjoint representation. As a result, in addition to the
\textquotedblleft Higgs" mode excluded earlier by the orthogonality
condition (\ref{a2}), all the other scalar fields are eliminated.
Consequently only the pure vector fields, $\mathcal{A}_{\mu ^{\prime }}^{ij}$
($\mu ^{\prime }=1,2,3$ ) or $\mathcal{A}_{\mu ^{\prime \prime }}^{ij}$ ($%
\mu ^{\prime \prime }=0,1,2$), for the time-like or space-like SLIV
respectively, are left in the theory. Clearly, the components $\mathcal{A}%
_{\mu ^{\prime }}^{(ij)=\left\langle ij\right\rangle }$ and $\mathcal{A}%
_{\mu ^{\prime \prime }}^{(ij)=\left\langle ij\right\rangle }$ correspond to
the true Goldstone vector boson, for each type of SLIV, respectively, while
all the other five ones (with $(ij)\neq \left\langle ij\right\rangle $) are
vector PGBs. Consequently these six modes altogether represent the
fundamental spin-connection field multiplet in the PGG\ theory in the final
symmetry broken phase.

\subsection{ Broken symmetry phase: zero tetrad modes}

Let us now turn to the tetrad fields. Again, as one can readily confirm, the
tetrad length-fixing constraint in (\ref{t1}) possesses the high total
global symmetry $ISO(4,12)_{WS}$ rather than$\ ISO(1,3)_{WS}\times
SO(1,3)_{LF}$ as other terms in the emergent PGG Lagrangian (\ref{l'}). This
symmetry then spontaneously breaks to some its "diagonal" subgroup $ISO(1,3)$
that results in an appearance of the corresponding Goldstone and Higgs
modes. Note that this violation actually appears as the 16-dimensional
Poincar\'{e} symmetry violation down to the ordinary 4-dimensional one. As
it is well known for spontaneously broken spacetime symmetries \cite{low},
such a violation can solely lead to the Goldstone modes corresponding to the
broken translational generators. There are no additional modes corresponding
to the broken Lorentz generators. So, we eventually have only twelve
Goldstone modes (according to the number of the broken translation
generators) which may be given by the non-diagonal $e_{\mu }^{i}$ components
($e_{1,2,3}^{0},$ $e_{2,3}^{1}$, $e_{3}^{2}$ and their inverse ones) ,
whereas the Higgs mode by some combination of their diagonal ones ($%
e_{0}^{0},$ $e_{1}^{1}$, $e_{2}^{2},$ $e_{3}^{3}$). Indeed, the above
Goldstone modes are in fact pseudo-Goldstone modes since, as was mentioned
above, the symmetry of the PGG Lagrangian $\mathcal{L}_{PGG}^{\mathfrak{em}}$
is much lower than the symmetry of the tetrad field constraint (\ref{t1}).

All that can be readily seen by using the familiar parametrization 
\begin{equation}
e_{\mu }^{i}\,=\mathfrak{e}_{\mu }^{i}\,+\mathrm{n}_{\mu }^{i}\sqrt{%
M_{e}^{2}-\mathfrak{e}^{2}}\text{\ \ }(\mathfrak{e}^{2}\equiv \mathfrak{e}%
_{\mu }^{i}\mathfrak{e}_{i}^{\mu }/\mathrm{n}^{2})  \label{par1}
\end{equation}%
with $\mathfrak{e}_{\mu }^{i}$ appearing as the vector Goldstone fields
which correspond to the spontaneous violation of the high-dimensional
translation invariance. For the unit vacuum direction tensors chosen
accordingly as $\mathrm{n}_{\mu }^{i}=\mathrm{\delta }_{\mu }^{i}$ and $%
\mathrm{n}_{i}^{\mu }=\mathrm{\delta }_{i}^{\mu }$ one therefore has 
\begin{equation}
\mathrm{\delta }_{\mu }^{i}\mathfrak{e}_{i}^{\mu }=0\text{ , \ }\mathrm{%
\delta }_{i}^{\mu }\mathfrak{e}_{\mu }^{i}=0\text{ \ (}\mathrm{\delta }_{\mu
}^{i}\mathrm{\delta }_{j}^{\mu }=\delta _{j}^{i},\text{ }\mathrm{\delta }%
_{\mu }^{i}\mathrm{\delta }_{i}^{\nu }=\delta _{\mu }^{\nu },\text{ }\mathrm{%
\delta }_{\mu }^{i}\mathrm{\delta }_{i}^{\mu }=4\text{).}  \label{gb}
\end{equation}%
At the same time, the vector Goldstone fields $\mathfrak{e}_{\mu }^{i}$ and $%
\mathfrak{e}_{i}^{\mu }$ fields are turn out to be the gauge fields of local
translations, as directly follows from the tetrad transformation law in (\ref%
{A}). Meanwhile the second (diagonal) term in the parametrization (\ref{par1}%
)\ represents the effective Higgs mode%
\begin{equation}
\mathfrak{h}=\sqrt{M_{e}^{2}-\mathfrak{e}^{2}}=M_{e}-\frac{\mathfrak{e}^{2}}{%
2M_{e}}+\cdot \cdot \cdot  \label{hm1}
\end{equation}%
Note that with this "mixed" Kronecker symbols $\mathrm{\delta }_{\mu }^{i}$
and $\mathrm{\delta }_{i}^{\mu }$\ one also has some new orthogonality
equation%
\begin{equation}
e_{k}^{\mu }e_{i}^{\nu }\mathrm{\delta }_{\nu }^{k}=e_{k}^{\mu
}e_{i}^{k}=e_{k}^{\mu }e_{\nu }^{k}\mathrm{\delta }_{i}^{\nu }=M_{e}^{2}%
\mathrm{\delta }_{i}^{\mu }  \label{t3}
\end{equation}%
provided that the standard orthogonality conditions (\ref{t1}) work\footnote{%
Actually, there are more orthogonality equations for all possible tetrad
components%
\begin{equation*}
e_{\rho }^{\nu }e_{\nu }^{\sigma }=\mathrm{\delta }_{i}^{\nu }e_{\rho }^{i}%
\mathrm{\delta }_{\nu }^{j}e_{j}^{\sigma }=\mathfrak{M}_{e}^{2}\delta _{\rho
}^{\sigma },\text{ }e_{k}^{i}e_{j}^{k}=\mathrm{\delta }_{\rho
}^{i}e_{k}^{\rho }\mathrm{\delta }_{j}^{\sigma }e_{\sigma }^{k}=\mathfrak{M}%
_{e}^{2}\delta _{j}^{i},\text{ }e_{k}^{i}e_{\rho }^{k}=\mathrm{\delta }%
_{\sigma }^{i}e_{k}^{\sigma }e_{\rho }^{k}=\mathfrak{M}_{e}^{2}\mathrm{%
\delta }_{\rho }^{i}
\end{equation*}%
in addition to the standard equations (\ref{t1}).}. Substituting the
parametrizations (\ref{par1}) into all them one can readily receive the
constraints put on the Goldstone fields $\mathfrak{e}_{\mu }^{i}$ and $%
\mathfrak{e}_{i}^{\mu }$ 
\begin{eqnarray}
\text{ }\mathfrak{e}_{\mu }^{i}\mathfrak{e}_{i}^{\nu }-\delta _{\mu }^{\nu }%
\mathfrak{e}^{2}+(\mathrm{\delta }_{\mu }^{i}\mathfrak{e}_{i}^{\nu }+\mathrm{%
\delta }_{i}^{\nu }\mathfrak{e}_{\mu }^{i})\mathfrak{h} &=&0,  \notag \\
\mathfrak{e}_{\mu }^{i}\mathfrak{e}_{j}^{\mu }-\delta _{j}^{i}\mathfrak{e}%
^{2}+(\mathrm{\delta }_{\mu }^{i}\mathfrak{e}_{j}^{\mu }+\mathrm{\delta }%
_{j}^{\mu }\mathfrak{e}_{\mu }^{i})\mathfrak{h} &=&0,  \notag \\
\text{\ }\mathfrak{e}_{\mu }^{i}\mathfrak{e}_{\nu }^{j}\,\mathrm{\delta }%
_{j}^{\mu }-\mathrm{\delta }_{\nu }^{i}\mathfrak{e}^{2}+2\mathfrak{e}_{\nu
}^{i}\mathfrak{h} &=&0.  \label{cs}
\end{eqnarray}

For a general metric tensor $g_{\mu \nu }(x)$ which corresponds to the
tetrad $e_{\mu }^{i}$ one consequently has from a conventional metric
definition (\ref{t2}) and equations (\ref{par1}) 
\begin{equation}
g_{\mu \nu }=\eta _{\mu \nu }+\frac{1}{M_{e}^{2}}[\mathfrak{h}(\mathrm{%
\delta }_{\mu }^{i}\mathfrak{e}_{i\nu }+\mathrm{\delta }_{\nu }^{j}\mathfrak{%
e}_{j\mu })+\mathfrak{e}_{\mu }^{i}\mathfrak{e}_{i\nu }-\eta _{\mu \nu }%
\mathfrak{e}^{2}]  \label{eta}
\end{equation}%
where $\eta _{\mu \nu }$ stands for a flat metric $\eta _{\mu \nu }=\eta
_{ij}\mathrm{\delta }_{\mu }^{i}\mathrm{\delta }_{\nu }^{j}$ in the world
space and, therefore, the second term in (\ref{eta}) represents a deviation
from the flat metric. As is readily seen from (\ref{eta}), the vacuum in the
PGG theory is a largely flat Minkowski spacetime that allows to treat
gravity as a generically spontaneously broken theory. Though this point was
discussed in many different contexts (see, for example, the review \cite{sar}%
), it looks the most transparent just in the emergent PGG framework. Indeed,
one can readily see that the above-mentioned deviation from a flat metric is
naturally small once the symmetry breaking scale $M_{e}$\ related to the
tetrad field $e_{\mu }^{i}$ is associated with the Planck mass scale $M_{P}$%
. Respectively, an inverse metric tensor $g^{\mu \nu }(x)$ corresponding to
the tetrad $e_{i}^{\mu }$ has a similar form with an extremely small
deviation from a flat metric $\eta ^{\mu \nu }=\eta ^{ij}\mathrm{\delta }%
_{i}^{\mu }\mathrm{\delta }_{j}^{\nu }$ given as in (\ref{eta}) by an
appropriate Goldstone tetrad field combinations. Indeed, due the constraints
(\ref{cs}), a conventional relationship between general metrics, $g_{\mu \nu
}g^{\nu \rho }=\delta _{\mu }^{\rho }$, is automatically satisfied.

\section{Emergent Einstein-Cartan theory and beyond}

\subsection{Non-propagating tetrads and spin-connections}

We start with the minimal theory part $\mathcal{L}^{(1)}$ (\ref{ll}) in the
basic emergent Lagrangian (\ref{larg}). Without kinetic terms, the tetrad
and spin-connection Goldstone modes in this minimal Lagrangian are not
propagating physical fields, though their variations may lead to some
non-trivial constraint equations. We will see below that, varying this
Lagrangian under Goldstone tetrad modes $\mathfrak{e}_{\mu }^{\;i}$ one
comes to the Einstein-Cartan equation, while variation under Goldstone
spin-connection modes $\mathcal{A}_{\mu }^{ij}$ may reveal some spin-spin
gravitational interaction trace in this equation.

Let us note first that for a variation of tetrad fields and their
determinant we have now taking into account that tetrads are dimensionful
fields, 
\begin{equation}
\delta e_{i}^{\mu }=-e_{i}^{\nu }e_{j}^{\mu }\delta e_{\nu }^{j}/M_{e}^{2},%
\text{ }\delta e=ee_{i}^{\mu }\delta e_{\mu }^{i}/M_{e}^{2}\text{ .}
\label{e-}
\end{equation}%
Multiplying the both sides of\ the first equation by $\mathrm{\delta }_{\mu
}^{i}$ and using the tetrad orthogonality condition (\ref{t3}) in its right
side one has%
\begin{equation}
\delta (\mathrm{\delta }_{\mu }^{i}e_{i}^{\mu })=-\delta (\mathrm{\delta }%
_{i}^{\mu }e_{\mu }^{i})
\end{equation}%
that due to the Goldstone condition (\ref{gb}) for the $\mathfrak{e}%
_{i}^{\mu }$ and $\mathfrak{e}_{\mu }^{i}$ modes, respectively, eventually
gives%
\begin{equation}
\delta \mathfrak{h}=0,\text{ }\delta e_{\mu }^{i}=\delta \mathfrak{e}_{\mu
}^{i}.
\end{equation}%
Thus, the effective Higgs field $\mathfrak{h}$ does not vary and a total
variation of the starting tetrad fields $e_{\mu }^{i}$($e_{i}^{\mu }$)
amounts to the variation of the pure Goldstone modes $\mathfrak{e}_{\mu
}^{i} $($\mathfrak{e}_{i}^{\mu }$). In terms of these modes the variation
equations (\ref{e-})\ acquire the simple forms 
\begin{equation}
\delta e_{i}^{\mu }=\delta \mathfrak{e}_{i}^{\mu }=-e_{i}^{\nu }e_{j}^{\mu
}\delta \mathfrak{e}_{\nu }^{j}/M_{e}^{2},\text{ }\delta e=\delta \mathfrak{e%
}=ee_{i}^{\mu }\delta \mathfrak{e}_{\mu }^{i}/M_{e}^{2}\text{ .}  \label{e=}
\end{equation}%
This in turn means that the variation of the minimal Lagrangian $\mathcal{L}%
^{(1)}$ (\ref{ll}) under the Goldstone tetrad fields $\mathfrak{e}_{\mu
}^{\;i}$ will lead to the same equations of motion as the variation under
the total tetrad fields $e_{\mu }^{\;i}$.

In contrast to tetrads, there is no the similar orthogonality conditions (%
\ref{t1}, \ref{t3}) for spin-connection fields $A_{\mu }^{ij}$. As a result,
not only its Goldstone mode $\mathcal{A}_{\mu }^{ij}$ but also its effective
Higgs mode $\mathcal{H}$ (\ref{hm}) will vary%
\begin{eqnarray}
\delta A_{\mu }^{ij} &=&\delta \mathcal{A}_{ij}^{\mu }+n_{ij}^{\mu }\delta 
\mathcal{H}  \notag \\
&=&\delta \mathcal{A}_{\mu }^{ij}-n_{\mu }^{ij}n^{2}\frac{\delta \mathcal{A}%
^{2}}{2\mathcal{H}}\simeq \delta \mathcal{A}_{\mu }^{ij}-n_{\mu }^{ij}n^{2}%
\frac{\delta \mathcal{A}^{2}}{2M_{A}}  \label{cor}
\end{eqnarray}%
that, therefore, might lead to the corrections of the order $\mathcal{O}(%
\mathcal{A}^{2}/M_{A}^{2})$ to the spin-connection constraint equation along
the vacuum direction given by the unit tensor $n_{ij}^{\mu }$. However, as
we show below, all these corrections are unavoidably cancelled in the final
Einstein-Cartan equation.

With these preliminary comments, let us now rewrite the minimal PGG theory $%
\mathcal{L}^{(1)}$ (\ref{ll}) in the symmetry broken phase. Indeed,
substituting the spin-connection field parameterization (\ref{a2}), one is
led to the Einstein-Cartan theory expressed in terms of the pure emergent
modes $\mathcal{A}_{\mu }^{ij}$. At the same time, one can still keep the
total tetrad field $e_{\mu }^{\;i}$ in the theory (being properly
dimensioned by mass scale $M_{e}$) rather than its Goldstone modes $%
\mathfrak{e}_{\mu }^{\;i}$ since , as was mentioned above, they both lead to
the same equations of motion in the minimal theory. However, as in the
Yang-Mills theory case considered in Section II B, one should first use the
local invariance of the emergent Lagrangian $\mathcal{L}^{\mathfrak{em}}$ (%
\ref{larg}) to gauge away the apparently large but fictitious Lorentz
violating terms (being proportional to the scale $M_{A}$) which appear in
the symmetry broken phase (\ref{a2}). As one can readily see, they stem from
the effective Higgs field $\mathcal{H}$ expansion (\ref{hm}) when it is
applied to some spin-connection field couplings following from the
corresponding covariant derivatives in the Lagrangian $\mathcal{L}^{%
\mathfrak{em}}$ . To exclude them we can make some appropriate Lorentz
rotations of all the fields involved, namely, spin-connection fields%
\begin{equation}
\mathcal{A}_{\mu \text{\ }}^{ij}\rightarrow \mathcal{A}_{\mu \text{\ }%
}^{ij}+\varepsilon _{k}^{i}\,\mathcal{A}_{\mu }^{kj}\,+\varepsilon _{k}^{j}\,%
\mathcal{A}_{\mu }^{ik}  \label{aa}
\end{equation}%
tetrads%
\begin{equation}
\text{\ }e_{\mu }^{i}\,\rightarrow e_{\mu }^{i}-\varepsilon _{k}^{i}\,e_{\mu
}^{k}  \label{bb}
\end{equation}%
and matter fermions%
\begin{equation}
\psi \rightarrow \left( 1+\frac{1}{4}\varepsilon ^{ij}\gamma _{ij}\right)
\psi   \label{pp}
\end{equation}%
with a phase $\varepsilon ^{ij}(x)$ being linear function in the
4-coordinate, $\varepsilon ^{ij}\,=-(n_{\mu }^{ij}x^{\mu })M_{A}$. These
transformations lead to an exact cancellation of the large constant term in
the effective Higgs field $\mathcal{H}$ expansion (\ref{hm}) so that the
transformed Lagrangian appears to contain everywhere just the combination $%
\mathcal{H}-M_{A}$ as an effective Higgs field\footnote{%
As in the non-Abelian theory case discussed in the footnote$^{8}$, the
vacuum shift of a spin-connection field multiplet $A_{\mu }^{ij}$ given in (%
\ref{a2}, \ref{hm}) being accompanied by a subsequent "rotation" (\ref{aa})
of an emergent multiplet $\mathcal{A}_{\mu \text{\ }}^{ij}$ can be written
entirely as%
\begin{equation*}
A_{\mu }^{ij}=\varepsilon _{k}^{i}\,\mathcal{A}_{\mu }^{kj}\,+\varepsilon
_{k}^{j}\,\mathcal{A}_{_{\mu }}^{ik}\,-\partial _{\mu }\varepsilon
^{ij}+n_{\mu \text{\ }}^{ij}(\mathcal{H}-M_{A}).
\end{equation*}%
with the transformation phase linear in the 4-coordinate. As can be readily
seen, the first two terms here present a pure gauge transfomation of the
spin-connection field multiplet $\mathcal{A}_{\mu }^{ij}$. Therefore, only
the third term can not be gauged away. Consequently, the transformed
Lagrangian contains just the combintion $\mathcal{H}-M_{A}$, as its
effective Higgs mode.}. Thus, the emergent Einstein-Cartan theory following
from the minimal Lagrangian (\ref{ll}) in the symmetry broken phase takes
the form (we retain the same notations for fields)

\begin{eqnarray}
e^{-1}\mathcal{L}_{EC}^{\mathfrak{em}} &=&\frac{1}{2\kappa }\frac{e_{i}^{\mu
}e_{j}^{\nu }}{M_{e}^{2}}\left[ \mathcal{R}_{\mu \nu }^{ij}+\overline{%
\mathcal{R}}_{\mu \nu }^{ij}(\mathcal{H}-M_{A})\right] +\frac{1}{2}\delta
(n^{\mu }\mathcal{A}_{\mu }^{ij})^{2}  \notag \\
&&+\frac{e_{i}{}^{\mu }}{2M_{e}}\left\{ \bar{\psi}\gamma ^{i}(%
\overleftrightarrow{\mathcal{D}}_{\mu }\psi )+\frac{1}{4}n_{\mu }^{ab}(%
\mathcal{H}-M_{A})\bar{\psi}[\gamma ^{i},\gamma _{ab}]\psi \right\}
\label{em3}
\end{eqnarray}%
where $\mathcal{R}_{\mu \nu }^{ij}$ is the stress tensor of emergent
spin-connection modes $\mathcal{A}_{\mu }^{ij}$%
\begin{equation}
\mathcal{R}_{\mu \nu }^{ij}=\partial _{\nu }\mathcal{A}_{\mu }^{ij}-\partial
_{\mu }\mathcal{A}_{\nu }^{ij}+\eta _{kl}(\mathcal{A}_{\nu }^{ik}\mathcal{A}%
_{\mu }^{lj}-\mathcal{A}_{\mu }^{ik}\mathcal{A}_{\nu }^{lj})  \label{r}
\end{equation}%
while $\overline{\mathcal{R}}_{\mu \nu }^{ij}$ stands for the new SLIV
oriented tensor of the type%
\begin{equation}
\overline{\mathcal{R}}_{\mu \nu }^{ij}=n_{\mu }^{ij}\partial _{\nu }-n_{\nu
}^{ij}\partial _{\mu }+\eta _{kl}\left[ (n_{\nu }^{ik}\mathcal{A}_{\mu
}^{lj}+n_{\mu }^{lj}\mathcal{A}_{\nu }^{ik}\mathfrak{)}-(n_{\mu }^{ik}%
\mathcal{A}_{\nu }^{lj}+n_{\nu }^{lj}\mathcal{A}_{\mu }^{ik})\right]
\label{r'}
\end{equation}%
acting on the effective Higgs field expansion terms in (\ref{em3}). The
"standard" Lorentz covariant derivative $\overleftrightarrow{\mathcal{D}}%
_{\mu }$ for fermion $\psi $, though written in terms of the emergent $%
\mathcal{A}$ fields, is defined exactly as in (\ref{de}). We have also
introduced a general axial gauge fixing term for the entire $\mathcal{A}%
_{\mu }^{ij}$ multiplet to remove all scalar modes from the theory\footnote{%
Note that we have omitted the higher order term $\eta _{kl}(n_{\nu
}^{ik}n_{\mu }^{lj}-n_{\mu }^{ik}n_{\nu }^{lj})(\mathcal{H}-M_{A})^{2}$ in
the Lagrangian (\ref{em3}) since it automatically vanishes. Indeed, the
background unit tensor $n_{\mu }^{ij}$ may always be chosen to be nonzero
for only one world space component $\mu $.}. After variation of the
Lagrangian (\ref{em3}) under tetrad field one comes to some extended
equation of motion that can be written in the form 
\begin{eqnarray}
&&\mathcal{R}^{\rho \sigma }-g^{\rho \sigma }\mathcal{R}/2+\kappa \mathcal{%
\vartheta }^{\rho \sigma }  \label{ext} \\
&=&-[\left( \overline{\mathcal{R}}^{\rho \sigma }-g^{\rho \sigma }\overline{%
\mathcal{R}}/2\right) +\frac{\kappa }{8M_{e}}(g^{\rho \sigma }e_{i}{}^{\mu
}-g^{\rho \mu }e_{i}^{\sigma })n_{\mu }^{ab}\bar{\psi}[\gamma ^{i},\gamma
_{ab}]\psi ](\mathcal{H}-M\mathfrak{_{\mathcal{A}}})\text{.}  \notag
\end{eqnarray}%
when going from local to general frame. Here the left side presents the
standard Einstein-Cartan equation terms including the energy-momentum tensor 
\begin{equation}
\mathcal{\vartheta }^{\rho \sigma }=\frac{1}{2M_{e}}(g^{\rho \sigma
}e_{i}{}^{\mu }-g^{\mu \sigma }e_{i}^{\rho })\bar{\psi}\gamma ^{i}%
\overleftrightarrow{\mathcal{D}}_{\mu }\psi
\end{equation}%
expressed, however, in terms of the emergent $\mathcal{A}_{\mu }^{ij}$
modes, whereas the right side corresponds to the Lorentz breaking background
terms newly appeared. The $\mathcal{R}$ and $\overline{\mathcal{R}}$ tensors
are defined as usual%
\begin{equation}
(\mathcal{R},\overline{\mathcal{R}})^{\sigma \rho }=(\mathcal{R},\overline{%
\mathcal{R}})_{\mu \nu }^{ij}e_{i}^{\sigma }e_{j}^{\nu }g^{\mu \rho
}/M_{e}^{2},\text{ }(\mathcal{R},\overline{\mathcal{R}})=(\mathcal{R},%
\overline{\mathcal{R}})_{\mu \nu }^{ij}e_{i}^{\mu }e_{j}^{\nu }/M_{e}^{2}%
\text{ .}
\end{equation}

The theory is not yet fully determined until the constraint equations for
the spin connection modes $\mathcal{A}_{\mu }^{ij}$ are found. Varying the
Lagrangian (\ref{em3}) w.r.t. these modes one has such equations in the
symmetry broken phase%
\begin{eqnarray}
&&\mathcal{D}_{\mu }\left( e_{a}^{[\mu }e_{b}^{\rho ]}\right) +\mathcal{A}%
_{ab}^{\rho }\overline{\mathcal{D}}_{\nu }\left( \frac{n^{2}}{2\mathcal{H}}%
n_{\mu }^{ij}e_{i}^{[\mu }e_{j}^{\nu ]}\right) +(\mathcal{H}-M_{A})\left[
n_{\mu a}^{i}e_{i}^{[\mu }e_{b}^{\rho ]}+n_{\mu b}^{i}e_{a}^{[\mu
}e_{i}^{\rho ]}\right]   \notag \\
&=&-\frac{\kappa }{4}M_{e}\left( e^{\rho k}\bar{\psi}\{\gamma _{k},\gamma
_{ab}\}\psi -n^{2}\frac{\mathcal{A}_{ab}^{\rho }}{2\mathcal{H}}e{}^{\mu
k}n_{\mu }^{ij}\bar{\psi}\{\gamma _{k},\gamma _{ij}\}\psi \right)   \label{L}
\end{eqnarray}%
where the first covariant derivative term in (\ref{L}) 
\begin{equation}
(\mathcal{D}_{\mu }-e^{-1}\partial _{\mu })\left( e_{a}^{[\mu }e_{b}^{\rho
]}\right) =\mathcal{A}_{\mu b}^{i}e_{a}^{[\mu }e_{i}^{\rho ]}+\mathcal{A}%
_{\mu a}^{i}e_{i}^{[\mu }e_{b}^{\rho ]}  \label{e1}
\end{equation}%
is indeed the standard one, while the second one

\begin{equation}
(\overline{\mathcal{D}}_{\nu }-e^{-1}\partial _{\nu })\left( \frac{n^{2}}{2%
\mathcal{H}}n_{\mu }^{ij}e_{i}^{[\mu }e_{j}^{\nu ]}\right) =\frac{n^{2}}{2%
\mathcal{H}}\eta _{kl}{\large (}n{\large _{\mu }^{ik}}\mathcal{A}_{\nu
}^{lj}+n{\large _{\nu }^{lj}}\mathcal{A}{\large _{\mu }^{ik})}e_{i}^{[\mu
}e_{j}^{\nu ]}  \label{e2}
\end{equation}%
is related to spontaneous Lorentz violation and disappears when its scale $%
M_{A}$ goes to infinity. Expanding the effective Higgs field $\mathcal{H}$ (%
\ref{hm}) in (\ref{em3}), one comes to the highly nonlinear theory in terms
of the zero spin-connection modes $\mathcal{A}_{\mu }^{ij}$ which contains
some properly suppressed Lorentz violating couplings. The point is, however,
that all these terms are precisely cancelled in the basic equation of motion
(\ref{ext}) once the constraint equations (\ref{L})\ are used. Thus, one
eventually is led to the standard Einstein-Cartan equation terms given
solely by the left side of the equation (\ref{ext}).

Let us show first how this cancellation works for the largest extra terms in
the equation (\ref{ext}). These terms correspond to the case when the tetrad
fields take the constant background value, $e_{i}^{\mu }=\mathrm{\delta }%
_{\mu }^{i}M_{e}$ ($e=1$). In this approach, which allows to omit all the
tetrad derivative terms in the constraint equations (\ref{L}), and also
leaving in them only terms linear in spin-connection modes $\mathcal{A}_{\mu
}^{ij}$, one comes to the "zero-order" constraint equations 
\begin{equation}
\mathcal{A}_{\mu b}^{i}\mathrm{\delta }_{a}^{[\mu }\mathrm{\delta }%
_{i}^{\rho ]}+\mathcal{A}_{\mu a}^{i}\mathrm{\delta }_{i}^{[\mu }\mathrm{%
\delta }_{b}^{\rho ]}=-\frac{\kappa }{4}\left( \mathrm{\delta }^{\rho k}\bar{%
\psi}\{\gamma _{k},\gamma _{ab}\}\psi -n^{2}\frac{\mathcal{A}_{ab}^{\rho }}{%
2M_{A}}\mathrm{\delta }^{\mu k}n_{\mu }^{ij}\bar{\psi}\{\gamma _{k},\gamma
_{ij}\}\psi \right) \text{.}  \label{LL}
\end{equation}%
They consequently give the following solution for spin-connection fields
expressed in pure local frame Lorentzian components 
\begin{equation}
\mathcal{A}_{abc}=-\ \frac{\kappa }{4}\bar{\psi}\widehat{\gamma }_{abc}\psi
\left( 1+\frac{\kappa }{8M_{A}}n^{2}\widehat{n}^{ijk}\bar{\psi}\widehat{%
\gamma }_{ijk}\psi \right)  \label{ok}
\end{equation}%
where the combination of the $\ \gamma $-matrices $\widehat{\gamma }_{ijk}$
and the matrix $\widehat{n}^{ijk}$ are defined according to the following
(anti)symmetrization of indices%
\begin{equation}
\widehat{\gamma }_{ijk}\equiv \frac{1}{2}\left( \gamma _{i[jk]}-\gamma
_{k[ij]}+\gamma _{j[ki]}+\eta _{ik}\eta ^{ab}\gamma _{a[bj]}+\eta _{jk}\eta
^{ab}\gamma _{b[ia]}\right) ,\text{ }\gamma _{k[ij]}\equiv \{\gamma
_{k},\gamma _{ij}\}.  \label{ga}
\end{equation}%
Note that the "zero-order" solution (\ref{ok}) holds in fact for the
contortion tensor $\mathcal{K}_{jab}$ part in the total spin-connection
field $\mathcal{A}_{abc}=\mathcal{A}_{abc}^{0}+\mathcal{K}_{abc}$ since an
ordinary part $\mathcal{A}_{abc}^{0}$ vanishes in the absence of the fermion
source. Putting this solution into the equation of motion (\ref{ext}) taken
in the same approximation (the background value for tetrads, no tetrad
derivative terms, no terms higher than linear in $\mathcal{A}_{\mu }^{ij}$)
one receives for the right side the vanishing sum of contributions 
\begin{equation}
\pm \frac{e\kappa ^{2}}{128M_{A}}\widehat{n}^{ijk}\left( \bar{\psi}\widehat{%
\gamma }_{ijk}\psi \right) \left( \bar{\psi}\widehat{\gamma }_{abc}\psi
\right) (\bar{\psi}\widehat{\gamma }^{abc}\psi )\text{.}  \label{orr}
\end{equation}%
stemming from its first and second terms , respectively. Thereby, one
unavoidably comes to the standard Einstein-Cartan equation in (\ref{ext}).
Though these six-fermion contributions are much smaller even regarding to
the standard four-fermion interaction in the Einstein-Cartan theory, their
cancellation is strictly provided by the Poincar\'{e} gauge invariance
emerged. Aplying these arguments order by order in spin-connection modes $%
\mathcal{A}_{\mu }^{ij}$ one may come to the same conclusion in a general
case as well.

\subsection{Theories with physical tetrad and spin-connection fields}

After the minimal Einstein-Cartant theory, let us now turn to a general PGG
Lagrangian (\ref{larg}) containing in its second part $\mathcal{L}^{(2)}$
all possible quadratic combinations of the Poincar\'{e} torsion and
curvature, $T_{\mu \nu }^{i}$ and $R_{\mu \nu }^{ij}$ (\ref{A}),
respectively. Substituting the field parametrizations (\ref{a2}) and (\ref%
{par1}) into the quadratic Lagrangian part in (\ref{larg}) and expanding the
square roots there in powers of $\mathcal{A}^{2}/M_{A}^{2}$ and $\mathfrak{e}%
^{2}/M_{e}^{2}$ we come, as in the above minimal model, to a highly
nonlinear theory in terms of the propagating tetrad and spin-connection
emergent Goldstone modes, $\mathcal{A}_{\mu }^{ij}$ and $\mathfrak{e}_{\mu
}^{i}$. Apart from the standard vector field couplings, this theory contains
many Lorentz and translation violating couplings stemming from their field
strengths $T_{\mu \nu }^{i}$ and $R_{\mu \nu }^{ij}$ in the symmetry broken
phase.

For some particular Lagrangian with propagating tetrad and spin-connection
fields due to the pure "Yang-Mills type" extension 
\begin{equation}
\mathcal{L}^{\mathfrak{em}}(\mathfrak{e},\mathcal{A},\psi )=\mathcal{L}%
_{EC}^{\mathfrak{em}}-\frac{e}{4\kappa _{e}}T_{\mu \nu }^{i}T_{i}^{\mu \nu }-%
\frac{e}{4\kappa _{A}}R_{\mu \nu }^{ij}R_{ij}^{\mu \nu }  \label{g}
\end{equation}%
one has applying the expressions (\ref{a2}, \ref{hm}, \ref{par1}) and
properly redefining fields according to the transformations (\ref{aa}, \ref%
{bb}, \ref{pp}) 
\begin{equation}
T_{\mu \nu }^{i}=\mathcal{T}_{\mu \nu }^{i}+\overline{\mathcal{T}}_{\mu \nu
}^{i},\text{ }R_{\mu \nu }^{ij}=\mathcal{R}_{\mu \nu }^{ij}+\overline{%
\mathcal{R}}_{\mu \nu }^{ij}(\mathcal{H}-M_{A})\text{ .}  \label{rr}
\end{equation}%
Here $\mathcal{T}_{\mu \nu }^{i}$ is an ordinary tetrad field stress tensor
expressed, however, in terms of emergent modes $\mathcal{A}_{\mu }^{ij}$ and 
$\mathfrak{e}_{\mu }^{i}$%
\begin{equation}
\mathcal{T}_{\mu \nu }^{i}=\partial _{\nu }\mathfrak{e}_{\mu
}^{i}\,-\partial _{\mu }\mathfrak{e}_{\nu }^{i}+\eta _{kl}(\mathcal{A}_{\nu
}^{ik}\mathfrak{e}_{\mu }^{l}-\mathcal{A}_{\mu }^{ik}\mathfrak{e}_{\nu
}^{l}),  \label{ti}
\end{equation}%
while $\overline{\mathcal{T}}_{\mu \nu }^{i}$ stands for a new tetrad
strength appearing in the symmetry broken phase 
\begin{eqnarray}
\overline{\mathcal{T}}_{\mu \nu }^{i} &=&[\mathrm{\delta }_{\mu
}^{i}\partial _{\nu }-\mathrm{\delta }_{\nu }^{i}\partial _{\mu }+\eta _{kl}(%
\mathcal{A}_{\nu }^{ik}\mathrm{\delta }_{\mu }^{l}-\mathcal{A}_{\mu }^{ik}%
\mathrm{\delta }_{\nu }^{l})]\mathfrak{h}  \notag \\
&&+\eta _{kl}\left[ n_{\nu }^{ik}\mathfrak{e}_{\mu }^{l}-n_{\mu }^{ik}%
\mathfrak{e}_{\nu }^{l}+(n_{\nu }^{ik}\mathrm{\delta }_{\mu }^{l}-n_{\mu
}^{ik}\mathrm{\delta }_{\nu }^{l})\mathfrak{h}\right] (\mathcal{H}-M_{A})\ 
\text{.}  \label{tj}
\end{eqnarray}%
The spin-connection stress tensors $\mathcal{R}_{\mu \nu }^{ij}$ and $%
\overline{\mathcal{R}}_{\mu \nu }^{ij}$ have already been given in (\ref{r})
and (\ref{r'}), respectively.

We see that the starting tetrad fields $e_{\mu }^{i}$ in the Lagrangian (\ref%
{g}) plays in fact the role of the Higgs multiplet for the spin-connection
fields $\mathcal{A}_{\mu }^{ij}$ due to which a part of them gets mass terms
of the type%
\begin{equation}
-\frac{M_{e}^{2}}{2k_{e}}(\mathcal{A}_{\mu }^{ij}\mathcal{A}_{ij}^{\mu }-%
\mathcal{A}_{\mu }^{ij}\mathcal{A}_{ik}^{\nu }\mathrm{\delta }_{\nu j}%
\mathrm{\delta }^{\mu k})
\end{equation}%
One can readily confirm that, similar to the conventional vector field
theories \cite{jej, cj}, the couplings related to the spin-connection fields
will not lead to physical Lorentz or translation violation effects. They
turn out again to be strictly cancelled among themselves in all processes
involved, like as the Compton scattering of $\mathcal{A}$ boson off the
matter fermion, the $\mathcal{A}-\mathcal{A}$ scattering and others.
Actually, their tree level amplitudes are essentially determined by an
interrelation between the longitudinal $\mathcal{A}$ boson exchange diagrams
and the corresponding contact $\mathcal{A}$ \ boson interaction diagrams
following from the higher terms in $\mathcal{A}^{2}/M_{A}^{2}$ in the
Lagrangian (\ref{g}). These two types of diagrams are always exactly
cancelled for any processes taken at least in the tree approximation.

The same can be said about the processes related to tetrad fields. In
contrast to the spin-connection couplings which are essentially similar to
those in the Yang-Mills theory \cite{jej}, tetrad couplings are somewhat
"hidden". Nevertheless, one can find from the Lagrangians (\ref{g}, \ref{em3}%
) that the lowest dimension ones are given by the operators%
\begin{eqnarray}
&&-\mathrm{\delta }_{\mu }^{i}(\partial ^{\mu }\mathfrak{e}_{i}^{\nu
})(\partial _{\nu }\mathfrak{e}^{2})/2M_{e},  \notag \\
&&\frac{\mathfrak{e}_{i}{}^{\mu }}{2M_{e}}\bar{\psi}\gamma ^{i}%
\overleftrightarrow{\partial }_{\mu }\psi ,\text{ }-\mathrm{\delta }%
_{i}^{\mu }\frac{\mathfrak{e}^{2}}{4M_{e}^{2}}\bar{\psi}\gamma ^{i}%
\overleftrightarrow{\partial }_{\mu }\psi \text{ ,}  \label{3c}
\end{eqnarray}%
where the first one presents the three-linear tetrad field coupling, while
the other two are related to the tetrad-fermion interactions. Again,
considering the scattering of tetrad $\mathfrak{e}$ field off the fermion $%
\psi $, one readily finds that their tree level amplitudes vanish being
provided by a mutual cancellation between the longitudinal $\mathfrak{e}$%
-boson exchange diagram which is determined by the first ($\mathfrak{e}^{3}$%
) and the second ($\mathfrak{e}\psi \psi $) vertices in (\ref{3c}), and the
corresponding contact $\mathfrak{e}$-boson interaction diagram ($\mathfrak{e}%
^{2}\psi \psi $). Likewise, it can be shown that all other symmetry breaking
processes appear unobservable.

Most likely, the same conclusion can be also derived for the Lorentz (or
translation) violation\ loop contributions of the spin-connection and tetrad
fields. However, the considered quadratic Lagrangian (\ref{g}) contains
ghosts and, therefore, should necessarily be complemented by some other
terms to exclude them from the theory. Fortunately, there exist the several
examples of the unitary PGG theories being free from ghosts and tachyons 
\cite{nev1, nev} where the above-mentioned loop calculations may appear
sensible. They include the theories with both torsion and curvature ($%
\mathcal{R}+\mathcal{R}^{2}+\mathcal{T}^{2}$), as well as the theories with
only torsion-squared ($\mathcal{R}+\mathcal{T}^{2}$)\ or curvature-squared ($%
\mathcal{R}+\mathcal{R}^{2}$) terms.

Some of these unitary PGG theories could be also used for an unification
with the Standard model. There is a point which can help to choose the right
PGG candidate. Actually, one may propose that the spin-connection fields $%
\mathcal{A}_{\mu }^{ij}$ could be unified with ordinary SM gauge fields in a
framework of some non-compact local symmetry group thus leading to a
hyperunification of all gauge forces presented in \ the local Lorentz frame.
As to tetrads $\mathfrak{e}_{\mu }^{i}$, however, they transform like as
ordinary matter fields being belonged to the fundamental vector multiplet of 
$SO(1,3)_{LF}$ rather than to its adjoint representation. Note that the
ordinary gauge theories do not contain the objects like the tetrads, namely,
the fundamental vector field multiplets. In this sense, one may only expect
a partial unification of PGG with SM unifying only the spin-connection
fields with the SM gauge bosons.

Remarkably, there is such an example of the unitary theory containing only
the curvature-squared terms \cite{nev1} that can be written in our notations
as%
\begin{equation}
\mathcal{L}^{\mathfrak{em}}(\mathcal{A},\psi )=\mathcal{L}_{EC}^{\mathfrak{em%
}}-\frac{e}{4\kappa _{A}}\mathcal{R}_{ijkl}\left( \mathcal{R}^{ijkl}+%
\mathcal{R}^{klij}-4\mathcal{R}^{ikjl}\right)  \label{re}
\end{equation}%
where the curvature tensors (\ref{r}) in the second term are properly
contracted with tetrads, $\mathcal{R}^{ijkl}=\mathcal{R}_{\mu \nu
}^{ij}e^{\mu k}e^{\nu l}$, and (anti)symmetrized. In this theory the tetrads
will only give the constraint equations causing some extra terms to the
Einstein-Cartan equation (\ref{ext}), whereas the spin-connection fields $%
\mathcal{A}_{\mu }^{ij}$ become to propagate like the gauge bosons in the
Standard Model. Their entire unification seems to be most obvious inside the
pseudo-orthogonal $SO(1,N)$ groups in which a direct embedding of the
Lorentz group $SO(1,3)$ can be readily carried out. Requiring then that such
unified group has to contain a non-trivial internal symmetry group giving
some grand unification theory (GUT) for three other forces, we come to the
condition $N-4=4k+2$ ($k=1,2,...$) selecting the $SO(N-4)$ GUTs which have
complex representations. So, the minimal possible symmetry group for a
hyperunification of all forces appears to be the $SO(1,13)$ which then
spontaneously breaks at some Planck mass order scale into $SO(1,3)\times
SO(10)$ so as to naturally lead to PGG, on the one hand, and $SO(10)$ GUT 
\cite{fr} for quarks and leptons, on the other\footnote{%
As we recently learned, such a type of the "GraviGUT" unification was first
discussed, while in the somewhat different context, quite a long ago \cite%
{per}.}. However, apart from the $SO(1,N)$ series, some other
hyperunification groups are also possible. Particularly, if one keeps an eye
on the $SU(N)$ type GUT, then the hyperunification groups may be looked for
in the special linear $SL(2N,C)$ groups containing as subgroups the $SL(2,C)$
covering the Lorentz group and some grand unified $SU(N)$ symmetry. Thus,
apart from a conventional $SU(5)$ GUT \cite{gg} which would stem from the
hyperunified $SL(10,C)$, the higher GUTs like as the $SU(8)$ \cite{ch, ro}
or $SU(11)$ \cite{ge} containing all three quark-lepton families could
emerge in turn from the hyperunified $SL(16,C)$ or $SL(22,C)$ theories,
respectively.

\section{Conclusion and overlook}

We have argued that the Poincar\'{e} gauge gravity could dynamically appear
in a general relativistic framework due to the covariant length-presering
constraints put on some prototype vector fields of spin-connections $A_{\mu
}^{ij}(x)$ and tetrads $e_{\mu }^{i}{(x)}$. Because of these constraints the
underlying global Poincar\'{e} symmetry in the theory converts into the
local one, thus leading to the PGG theory being gauged by these prototype
fields. The point is, however, that these gauge fields are turned out to be,
at the same time, the vector Goldstone and pseudo-Goldstone modes, $\mathcal{%
A}_{\mu }^{ij}$ and $\mathfrak{e}_{\mu }^{i}$, manifesting themseves in the
symmetry broken phase of the accidental global symmetries accompanying the
covariant constraints (\ref{t1}, \ref{a1}). As in the other emergent gauge
theories, the emergent Poincar\'{e} gauge invariance in PGG makes any
symmetry violation effects in the theory unobservable. In this connection,
the above constraints appear as the covariant gauge conditions being then
reduced to the noncovariant gauge choice in the accidental symmetry broken
phase. Thus, this global symmetry violation is only manifested as the loss
of gauge degrees of freedom of massless vector Goldstone bosons.

Another important point concerns the vector field mediation in PGG at high
energies which may appear through the fundamental spin-connection and tetrad
fields, $\mathcal{A}_{\mu }^{ij}$ and $\mathfrak{e}_{\mu }^{i}$ (together
with the low energy mediation through the effective tensor metric field $%
g_{\mu \nu }$). This makes the gauge gravity to be closer with other
interactions that could lead in principle to its unification with other
basic gauge forces provided by the Standard Model. Such hyperunification may
presumably be related to a generic analogy between local frame in PGG and
internal charge space in conventional quantum field theories, as was
mentioned above. Actually, some extended non-compact internal space may
contain a local frame spacetime $M_{4\text{ }}$as its subspace. This looks
quite reverse to the well-known approaches \cite{isham, cho, lisi} when a
part of some extended spacetime appears as an internal space. In contrast,
now just geometry may follow from charges rather than charges from geometry%
\footnote{%
An interesting attempt to treat the global Poincar\'{e} symmetry as the pure
internal one has been made in \cite{weis}}. We presented above some possible
hyperunifying GUT candidates for such an unification of all four fundamental
forces.

One might expect that there would be a potential danger for any hyperunified
theory due to the Coleman-Mandula "no-go" theorem \cite{18} on the
impossibility of combining spacetime and internal symmetries. Nonetheless,
regarding to the hyperunified theories we consider here, this "no-go"
theorem seems not to be a unavoidable obstacle, as may be seen from the
following heuristic arguments. Indeed, the first is that the theorem only
works if there is a mass gap in the theory that means difference in energy
between the vacuum and the next lowest energy state which is in fact the
mass of the lightest particle. However, there is no mass gap in the unified
theory in the hyperunification symmetry limit where all fields, both gauge
bosons and matter fields, are massless. Apart from the extended gauge
invariance in the hyperunified theories, the generic masslessness of all the
gauge fields involved (like as photon, gluons and graviton) could also be
provided by their Nambu-Goldstone nature being related to the spontaneous
breakdown of global spacetime symmetries \cite{cfn,kraus,kos,car}. The
second and rather important point seems to be related to the nature of PGG
as a theory where a gauge group does not need to be specially linked to the
base space manifold. Actually, one may take the space to be either curved 
\cite{Utiyama} or flat \cite{Kibble} being no conditioned by PGG on its own.
As usually appears \cite{Kibble, mansouri1}, one would have to identify the
theory with some space manifold at a later stage. As a result, the local
frame Lorentz gauge symmetry rather looks like an internal symmetry in PGG,
and as such may then have an unobstructed unification with SM or GUT.

We will return to these interesting issues elsewhere \cite{chka}.

\section*{Acknowledgments}

I would like to thank Ian Darius, Colin Froggatt and Holger Nielsen, as well
as the colleagues and students at Center for Elementary Particle Physics of
Ilia State University, especially Zurab Kepuladze, Luka Megrelidze and Zurab
Tavartkiladze, for interesting discussions and useful remarks. This work is
partially supported by Georgian National Science Foundation (Contracts No.
31/89 and No. DI/12/6-200/13).

\end{document}